\documentclass[8pt]{article}

\usepackage[table]{xcolor}
\usepackage{graphicx,pslatex,float,color,array,amssymb,amsmath,euscript}
\usepackage{pxfonts}
\usepackage{ifthen}
\usepackage{longtable}
\usepackage{eso-pic}
\usepackage{sidecap}
\usepackage{xspace}
\usepackage[paperwidth=21.0cm,paperheight=29.7cm,textwidth=16cm,textheight=23cm,top=3cm,left=2.5cm]{geometry}
\definecolor{navy}{rgb}{0,0,0.5}
\usepackage[pdftex,
           hyperindex=true,
           colorlinks=true,
           linkcolor=navy,
           anchorcolor=magenta,
           citecolor=navy,
           urlcolor=navy,
           unicode,
           implicit=true]{hyperref}

\renewcommand{\vec}[1]{\boldsymbol{#1}}
\newcommand{\be}{\begin{equation}}
\newcommand{\ee}{\end{equation}}
\newcommand{\bes}{\begin{equation*}}
\newcommand{\ees}{\end{equation*}}

\newcommand{\ddp}[2]{\frac{\partial #1}{\partial #2}}

\newcommand{\amacro}{a_{\mbox{\tiny macro}}}
\newcommand{\amicro}{a_{\mbox{\tiny micro}}}
\newcommand{\Aasp}{A_{\mbox{\tiny asp}}}
\newcommand{\ddps}[2]{\frac{\partial^2{#1}}{\partial{#2}^2}}
\newcommand{\bmtwo}{\left[\begin{array}{cc}}
\newcommand{\bv}{\left[\begin{array}{c}}
\newcommand{\ema}{\end{array}\right]}
\renewcommand{\vec}[1]{\boldsymbol{#1}}

\title{The Elastic Contact of Rough Spheres\\ Investigated Using a Deterministic Multi-Asperity Model}
\author{Vladislav A. Yastrebov}
\date{\small{\it MINES ParisTech, PSL Research University, Centre des Mat\'eriaux\\ CNRS UMR 7633, BP 87, 91003, Evry, France}}

\begin{document}

\maketitle

\begin{flushleft}
 \Large{\bf Abstract.}\normalsize
\end{flushleft}

\noindent 
In this paper we use a deterministic multi-asperity model to investigate the elastic contact of rough spheres. Synthetic rough surfaces with controllable spectra were used to identify individual asperities, their locations and curvatures. The deterministic analysis enables to capture both particular deformation modes of individual rough surfaces and also statistical deformation regimes, which involve averaging over a big number of roughness realizations. Two regimes of contact area growth were identified: the Hertzian regime at light loads at the scale of a single asperity, and the linear regime at higher loads  involving multiple contacting asperities. The transition between the regimes occurs at the load which depends on the second and the fourth spectral moments. It is shown that at light indentation the radius of circumference delimiting the contact area is always considerably larger than Hertzian contact radius. Therefore, it suggests that there is no scale separation in contact problems at light loads. In particular, the geometrical shape cannot be considered separately from the surface roughness at least for approaching greater than one standard roughness deviation.

\begin{flushleft}
 {\bf Keywords.} roughness; indentation; contact; deterministic multi-asperity model; contact area.
\end{flushleft}

\section{Introduction}

Contact and friction interactions play an essential role in many quotidian contexts, including those related to industry and transportation (e.g., tire-road and wheel-rail contacts, electric switches, gears, bearings, and brake systems), everyday human activity (e.g., walking, handling, touching, and sitting) and natural phenomena (e.g., earthquakes, landslides, and glacier motion). Regardless of such prevalence, contact-related mechanisms (friction, adhesion, and wear) are still not fully understood and thus are among the most cutting-edge research topics in the mechanical community~\cite{mo2009friction,ben2010dynamics,pastewka2011anisotropic,aghababaei2016critical}.

Numerous models of contact-related mechanisms exist at structural scale~\cite{vakis2018}. 
They serve to model (1) interfacial normal and tangential stiffness~\cite{wriggers1993application,akarapu2011stiffness}, (2) frictional resistance~\cite{ruina1983slip}, (3) material removal on rubbing surfaces (wear)~\cite{archard1956wear,yin2010adhesive}, (4) heat transfer between contacting solids~\cite{madhusudana1996thermal}, (5) contact electric resistance~\cite{slade2017electrical,yastrebov2015holm}, (6) adhesion~\cite{fuller1975effect,pastewka2014contact}, (7) interfacial fluid flow~\cite{dapp2012self,dapp2016fluid,shvarts2018trapped}, (8) microstructural changes in near-contact material layers~\cite{ramesh2008modeling}, (9) fretting wear life-cycle~\cite{hills1994mechanics,fouvry1997wear,dick2006fretting}, (10) debris generation~\cite{sayles1982influence,basseville2011numerical},
(11) lubrication~\cite{patir1978average}, and other mechanisms. 
% % % % % % % % % % % % % % % % % % % % % % 
The associated models can be incorporated in a macroscopic/structural model via constitutive interfacial equations. These equations can be based either on experimental data, and thus remain purely phenomenological, or can take the microscopic roughness as the starting point. 
The latter class of models shall have a greater predictive power, and potentially can be used for a large spectrum of applications. However, because of the strong non-linearity of the contact/friction mechanisms and extreme complexity of surface roughness, construction of a reliable analytical micro-mechanical model presents a serious challenge. Moreover, the scale separation which would allow to separate the macroscopic shape of contacting solids and the accompanying roughness cannot be always ensured.

% % % % % % % % % % % % % % % % % % % % % % 
Admittedly, all the aforementioned phenomena are strongly related to the surface roughness.
(1) The compression of contacting asperities as well as their interlocking contributes to the interfacial stiffness. 
(2) According to the adhesive theory of friction the frictional resistance of contacting spots is pressure independent and thus the total frictional resistance of the interface is proportional to the true contact area. In general, this area may depend on the hold-time or sliding velocity manifesting itself as a velocity-dependent friction coefficient.
(3) In case of adhesive wear, the removed volume depends on the characteristics and number of contact spots forming the true contact area. 
(4) The overall thermal conductivity of the interface is determined by heat (phonons and, if available, electrons) passing directly through contact spots, by convective heat transfer through the interface fluid and also by the radiative heat transfer. All these heat exchange mechanisms depend on the true contact area and on the opening (or gap) field between non-contacting parts of the two solids. 
(5) In a rather similar way, the electric current passes through the true contact area between contacting solids subjected to a difference of the electric potential. However, due to the presence of non-conducting oxides on the surface the effective conducting area is, in general, smaller than the true contact area. 
(6) Adhesion between solids is also considerably controlled by roughness and the effective contact area. On the other hand, the adhesion exploited by insects and small animals such as gecko~\cite{scherge2001biological,huber2007influence} is more related to the hierarchical structure of their limbs. 
(7) Fluid flow between contacting (or slightly separated) surfaces in sealing applications is controlled by their roughnesses, the out-of-contact areas, forming an opening field, and enabling the fluid to pass through the interface. Under certain critical load no possible paths exist, which corresponds to the percolation point. 
(8) Frictional sliding on contacting asperities is a dissipative process, the produced heat partly goes into the rubbing solids and may results in huge thermal gradients and very high localized temperatures at sliding asperities~\cite{bowden_tabor,rice2006heating}. These extreme thermo-mechanical loads can induce phase transformations and changes in material microstructures.
(9) Local wear and initiation of fatigue cracks in fretting conditions~\cite{proudhon2005ijf,kubiak2011ti} strongly depend on the local stress state, which is intimately linked with the surface roughness.
(10) Produced by adhesive wear~\cite{sasada1984w}, debris particles are necessarily related to the true contact area as discussed in (3); geometrical characteristics of surfaces are less important for the debris formed in abrasive wear~\cite{hokkirigawa1988ti}. The presence of debris (or mobile third body) can significantly affect the true contact area and thus modify the kinetics of debris production.
(11) Effect of the true contact area on the boundary lubrication is straightforward, concerning the hydrodynamic lubrication regimes and mixed lubrication the opening (gap) field plays an essential role and determines the film thickness, its possible breakage and thus determines the resulting friction and wear.

In all the describing processes and phenomena the mechanical interaction between rough surfaces  plays the key role. 
Accurate prediction of this mechanical interaction would allow constructing reliable and physically sound models for all the aforementioned phenomena.
Mechanical models of rough contact can be divided into two main classes: deterministic and statistical models. The former models rely on direct numerical simulations of contact~\cite{pei2005jmps,campana2006practical,anciaux2009ijnme,yastrebov2011rough,putignano2012ijss,afferrante2012w,bemporad2015optimization,muser2017meeting} and take as an input the topographical data of contacting surfaces and their material properties in order to predict integral quantities such as interface stiffness,  evolution and growth of  true contact area and of the mean gap under increasing pressure, etc. In addition, such models allow to study the shape and spatial distribution of contacting clusters, local stress states, critical zones, etc. These models are especially important for situations when the contact area is small or when the roughness of contacting solids cannot be seen as homogeneous at the scale of contact area, for example, the rough surface can have critical defects, rare high asperities or rare third-body particles. 
On the contrary, the statistical models~\cite{greenwood1966prcl,bush1975w,persson2001jcp,carbone2008jmps} require only a few parameters of surface roughness (for example, root-mean squared height or surface gradient, probability density, fractal dimension) and are appropriate for situations of homogeneous roughness. 
These models can yield results in terms of  statistical distributions (for example, pressure and tangential tractions) and mean integral values (for example, mean spacing between contacting asperities, integral contact area).
Such models are, in general, easier to use than deterministic but on the other hand they are inherently less precise since cannot take into account finite size of the contact area and cannot accurately resolve complex mechanical interaction between contacting spots.

In this paper we consider a contact problem between two non-conformal solids with superposed roughness. This generalization of Hertzian contact is relevant to almost all aforementioned applications since in real engineering and natural systems, the contact can be usually reduced to a contact between two smooth surfaces, whose shape is described by a two-form. Here however, we limit our attention to the contact of solids of revolution. To address this problem which was already studied using different models in different regimes~\cite{greenwood1967elastic,goryacheva2006mechanics,goryacheva2013contact,pastewka2016contact}, we choose to use a deterministic multi-asperity approach, rather similar to ones elaborated in~\cite{afferrante2012interacting,afferrante2018elastic}, which includes long-range elastic interaction between asperities. The main objective is to demonstrate capacities of the model in predicting particular features of this contact interaction for a given (deterministic) roughness. On the other hand we will explore its ability to derive statistical results, which, however, can be also done using Greenwood and Tripp's model~\cite{greenwood1967elastic}. The deterministic multi-asperity model, compared to statistical multi-asperity models~\cite{greenwood1966prcl,bush1975w,carbone2008jmps}, takes into account long-range interactions; and it enables to handle accurately very light loads, which are hardly reachable for conventional direct numerical simulations using finite element~\cite{pei2005finite,yastrebov2011cras} or modifications of boundary element methods~\cite{stanley1997,polonsky1999numerical,campana2006practical,bemporad2015optimization}. 

The paper is organized as follows. In Section~\ref{sec:1} we define the problem, recall the roughness generation method and explain the method to extract asperities' data. In Section~\ref{sec:2}, the numerical model to solve contact problems is presented. The results are presented in Section~\ref{sec:3}, namely (1) the analysis of the true contact area evolution, the two relevant regimes and transition between them;  (2)  the study of contact pressure and contact area  as well as their spatial and statistical distributions. We conclude by a short discussion of main results in Section~\ref{sec:4}.

% \section{Statistical models}
% 
% Starting from Archard and GW models, further Persson
% 
% \section{Deterministic models}
% 
% FEM, BEM, GFMD, MD, Carbone with interactions, Durand with interactions

\section{\label{sec:1}Problem set-up and roughness}

\subsection{Set-up}

\begin{figure}
  \includegraphics[width=1\textwidth]{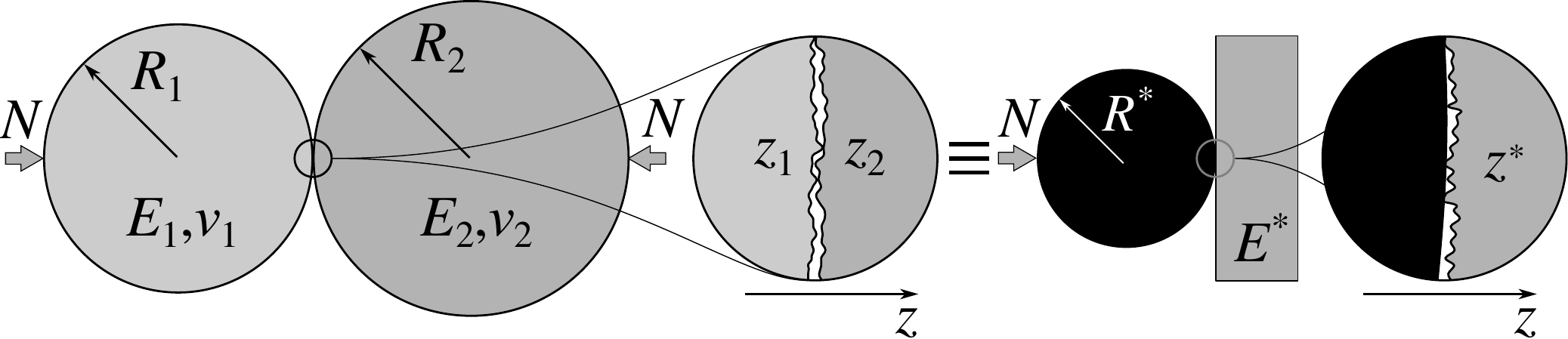}
 \caption{\label{fig:setup}Contact between two elastic spheres with rough surfaces and the equivalent problem of indentation of an effective elastic and rough half-space by a rigid smooth sphere of effective radius.}
\end{figure}

Using the deterministic multi-asperity model with long-range interaction we will consider a problem of contact between locally axisymmetric parabolic solids [see Fig.~\ref{fig:setup}(a)] of curvature radii $R_1$ and $R_2$, the solids are assumed to be elastic, homogeneous and isotropic with elastic properties $E_1,\nu_1$ and $E_2,\nu_2$, respectively, where $E$ and $\nu$ are the Young's modulus and Poisson's ratio, respectively. Schematically this problem can be presented as a contact between two spheres\footnote{We recall that in the vicinity of the tip, a sphere and a paraboloid of revolution of the same curvature are indistinguishable.}, see Fig.~\ref{fig:setup}. We assume that the contacting solids possess rough surfaces, which, in the close vicinity of contact zone, can be described by analytical functions $z_1(x,y)$ and $z_2(x,y)$. Note that this roughness in case of curved surfaces shall be superimposed on the macroscopic shapes, which can be assumed to be $z_{si}=(x^2+y^2)/(2R_i)$. We assume that  we can neglect the tangential (in-contact-plane) displacement of surface points, i.e. the effective roughness is given by $z^*(x,y) = z_1(x,y)-z_2(x,y)+z_0$, where $z_0$ is a constant. Under frictionless elastic contact between rough surfaces with relatively small slopes, this  problem can be mapped on a problem between a smooth rigid sphere with effective radius $R^* = R_1R_2/(R_1+R_2)$ and a rough elastic half-space with effective elastic modulus $E^*$~\cite{johnson1985b,barber2003bounds}:
$$E^* = \frac{E_1E_2}{(1-\nu_2^2)E_1 + (1-\nu_1^2)E_2}.$$
In case of smooth surfaces, we recall~\cite{johnson1985b} below the analytical solutions for this problem under the action of a squeezing force $N$. The contact radius $a$,  the approach between distance points $\delta$, and the pressure distribution $p(x,y)$ are given by the following expressions:
\begin{equation}
a = \left(\frac{3NR^*}{4E^*}\right)^{1/3},\quad \delta = \frac{a^2}{R^*} = \left(\frac{9N^2}{16R^*E^{*2}}\right)^{1/3},\quad p(x,y) = \frac{3P}{2\pi a^2} \sqrt{1-\frac{x^2+y^2}{a^2}}.
\end{equation}

\subsection{Roughness}

To model the surface roughness we use a white-noise filtering technique~\cite{hu1992ijmtm,yastrebov2017jmps}. This technique enables us to control accurately the spectrum of the surface and also to preserve the stochastic aspect of real roughness. Namely, we control (1) the Hurst exponent $H$, which determines the power-law decay (at least for short wavelengths $\lambda$) of spectral energy $\Phi \sim \lambda^{2(1+H)}$, and also (2) the lower and (3) the upper cut-off wavelengths $\lambda_s$ and $\lambda_l$, respectively, where lower indices $s$ and $l$ stand for ``short'' and ``long''. At the same time the spectrum can be made bi-fractal: a power-law increase with the wavelength in the interval $\lambda\in[\lambda_s,\lambda_l]$ can be followed by a plateau for longer wavelengths $\lambda\in(\lambda_l,L]$, where $L$ is the sample length.

Based on this definition, we can readily conclude that for loads implying the value of the macroscopic contact radius $\amacro$ which is comparable with the shortest roughness wavelength $\amacro\sim\lambda_s$, the contact will be formed by a few contact spots, whereas for $\amacro \gg \lambda_s$ it is expected that the contact contains a statistically meaningful number of spots. Therefore, statistical models could be accurately applied only for the later case. At the same time, it is known that the longest wavelength (of the power-law spectral decay) $\lambda_l$ determines the dispersion of results~\cite{yastrebov2012pre,pastewka2013pre,yastrebov2015ijss,rey2017hal}: therefore, to have a spatially homogeneous and statistically reproducible contact-spot distribution for all possible roughness realizations, the contact radius has to fulfil the following condition $\amacro \gg \lambda_l$ as well. However, this last condition should be fulfilled only for generated rough surfaces possessing discrete spectra.

Another important consideration is that in multi-asperity models it is assumed that local contact spots are located far from each other. A softer restriction would be the assumption that the contact zones associated with different asperities do not intersect. This strong limitation was partly overcame in~\cite{afferrante2012w} by assuming that coalescing contact spots (non-convex and possibly non-simply connected) form a new effective one of  a circular shape. However, this improvement requires to switch from statistical to a deterministic model. Moreover, separate asperities located on a smooth surface are inevitably separated by valleys and their merge is associated with a formation of junctions over saddle points~\cite{yastrebov2014tl,dapp2015epl}, which cannot be described in the framework of Hertzian theory which assumes the parabolic shape of contacting solids.
Rigorously in the framework of statistical model we shall assume that at least the contact radius associated with asperities does not overpass the quarter\footnote{It can be assumed that $z$-curvature of an in-plane line connecting two neighbouring asperities, changes at quarter distance from each summit.} of the average distance between asperities $\langle d\rangle$ (see~\ref{app} for details), i.e.
\begin{equation}
  \amicro < \frac{\langle d\rangle}{4} = \frac{1}{4\sqrt D} \approx \lambda_l \sqrt{\frac{\sqrt3 }{8  \pi } \frac{(2-H)}{(1-H)}\frac{(\zeta^{2-2H} - 1)}{\zeta^{4-2H}}} %\lambda_s\sqrt{\frac{\sqrt{3}(2-H)}{8\pi(1-H)}},
  \label{amicro_limit}
\end{equation}
where $D$ is the density of asperities, and the last approximate equality is valid for surfaces with a rich spectral content, i.e. for $\zeta = \lambda_l/\lambda_s \gg 1$.

In most statistical models, both multi-asperity and Persson's model, the rough surface is assumed to be Gaussian, i.e. the heights have an infinite support.
A notable exception is the study of asperity-height distribution on a semi-infinite support~\cite{ciavarella2017asme}. Therefore, for any reasonable approaching (here, the distance between mean planes of rough surfaces expressed in roughness-dispersion units), a small but non-zero portion of the surface will experience an infinite displacement, which would readily result in an apparent violation of criterion~\eqref{amicro_limit}. 
However, since in statistical models the notion of asperities is hidden behind the height probability density function, it seems reasonable to find the mean local asperity radius by assuming that the contact area $\Aasp$ predicted by a multi-asperity model over nominal area $A_0$ is formed by $N$ identical contact spots $\pi\amicro^2$, where the number of spots $N$ is assumed to be known through the asperity density $D$ as $N = DA_0$, therefore the individual contact-spot radius can be estimated as
\begin{equation}
  \amicro \approx \sqrt{\frac{\Aasp}{\pi D A_0}},
\end{equation}
where from using~\eqref{amicro_limit}, a simple limit for the validity of multi-asperity models could be established: 
\begin{equation}
 \frac{\Aasp}{A_0} < \frac{\pi}{16} \approx 19.6\%.
\end{equation}
This limit can be used locally (point-wise) as the upper bound for multi-asperity models, i.e. the local area fraction predicted by multi-asperity model cannot overpass $19.6$ \%. Of course, because of non-uniform distribution of asperities and contact spots, the real limit can be considerably smaller (see also comparison between deterministic and multi-asperity models~\cite{yastrebov2017jmps}).

\subsection{Asperities}

In vicinity of every summit, a smooth surface can be approximated by a quadratic form of local coordinates:
$$ 
  z(x,y) = z_0 + a x^2 + b y^2 + 2 c xy = z_0 + \bv x\\y\ema^\intercal \bmtwo a \;\;& c \\ c \;\; & b\ema \bv x\\y\ema = z_0 + \vec X^\intercal \vec A \vec X
%                                                 \end{bmatrix} \begin{bmatrix} x \\ y\end{bmatrix},
$$
with $$a = \frac12\ddps{z}{x}, \quad b = \frac12\ddps{z}{y}, \quad c = \frac12\frac{\partial^2 z}{\partial x\partial y}.$$
To determine the surface principal curvatures and their orientation, we seek for such rotation $\varphi$ of $\{x,y\}$ coordinates 
$$
  \vec X = \vec Q \vec X',\quad \vec X =  \bv x\\y\ema,\quad \vec X' =  \bv x'\\y'\ema,\quad \vec Q =  \bmtwo \cos(\varphi) & \sin(\varphi)\\ -\sin(\varphi) & \cos(\varphi) \ema
%    \bv x\\y\ema = 
$$
that the resulting form is diagonal:
$$
  z(x,y) = z_0 + \vec X'^\intercal \vec Q^\intercal \vec A \vec Q \vec X', \quad \vec Q^\intercal \vec A \vec Q = \bmtwo \lambda_1 \;\; & 0 \\ 0\;\; & \lambda_2 \ema.
$$
The angle $\varphi$ that ensures the diagonality should verify the following equation:
$$
  (a - b) \cos(\varphi)\sin(\varphi) + c (\cos^2(\varphi) - \sin^2(\varphi))  = 0,
$$
which after dividing by $(c\cos^2(\varphi))$ gives a quadratic equation for $\tan(\varphi)$:
$$
  \tan^2(\varphi) - \frac{a-b}{c}\tan(\varphi)  - 1 = 0,
$$
therefore the two solutions are given by :
\begin{equation}
 \varphi_{i} = \mathrm{arctan}\left(\frac{a-b}{2c} \pm \sqrt{\left(\frac{a-b}{2c}\right)^2+1}\,\right),
 \label{eq:angle}
\end{equation}
which correspond to eigenvalues 
$$
 \lambda_{i} = \frac12\left(a+b \pm \sqrt{(a-b)^2 + 4c^2}\right),
$$
therefore the principal curvatures and principal radii are given by:
\begin{equation}
  \kappa_{i} = \left(a+b \pm \sqrt{(a-b)^2 + 4c^2}\right),\quad R_{i} = 1\left/\left(a+b \pm \sqrt{(a-b)^2 + 4c^2}\right)\right.
  \label{eq:curvatures}
\end{equation}

Rough surface measurements are inevitably of discrete nature, therefore the question of asperity identification is relevant~\cite{paggi2010w}. 
Numerically generated rough surfaces~\cite{hu1992ijmtm,meakin1998fractals} (as soon as they are not given by analytical functions) are also discrete, i.e. the surface roughness is determined by a matrix of heights $z_{ij}$, where lower indices $i,j$ determine the in-plane position as $z_{ij} = z(i\Delta x, j\Delta y)$, where $\Delta x,\Delta y$ are the sampling intervals in $x,y$ directions, respectively. 
For such a representation, every potential summit $z'_{ij}$ is identified if the nearest and next-nearest (diagonal) neighbour points are lower:
$z'_{ij} > z\left[(k+i)\Delta x,(l+j)\Delta y\right]$, where $k,l$ may take values $\{-1,0,1\}$ except $\{k,l\}=\{0,0\}$. Using the finite differences, for all generalized summits the following values of the components of the fundamental matrix $\vec A$ can be estimated as:
\begin{equation}
  a = \frac12\ddp{^2\!z}{x^2} \approx \frac{z_{i+1j}+z_{i-1j}-2z_{ij}}{2\Delta x^2};\quad b=\frac12\ddp{^2\!z}{y^2} \approx\frac{z_{ij+1}+z_{ij-1}-2z_{ij}}{2\Delta y^2}
  \end{equation}
  \begin{equation}
  c = \frac12\ddp{^2\!z}{x\partial y} \approx \frac{z_{i+1j+1}+z_{i-1j-1}-z_{i+1j-1}-z_{i-1j+1}}{8\Delta x \Delta y}.
\end{equation}
The principal curvatures and radii then can be found from~\eqref{eq:curvatures}.
If both curvatures are negative the generalized summit is the real summit. If the principal curvatures have a different sign $\kappa_1\kappa_2 < 0$ this ``summit'' is a saddle point masked by discretization. 
The algebraic $\kappa_a$ and the geometric $\kappa_g$ mean curvatures are defined as follows
\begin{equation}
    \kappa_a = \frac12(\kappa_1+\kappa_2) = \frac12(a+b)
    \end{equation}
    \begin{equation}
    \kappa_g = \sqrt{\kappa_1\kappa_2} = 2\sqrt{ab - c^2}
\end{equation}

\begin{figure}
  \includegraphics[width=1\textwidth]{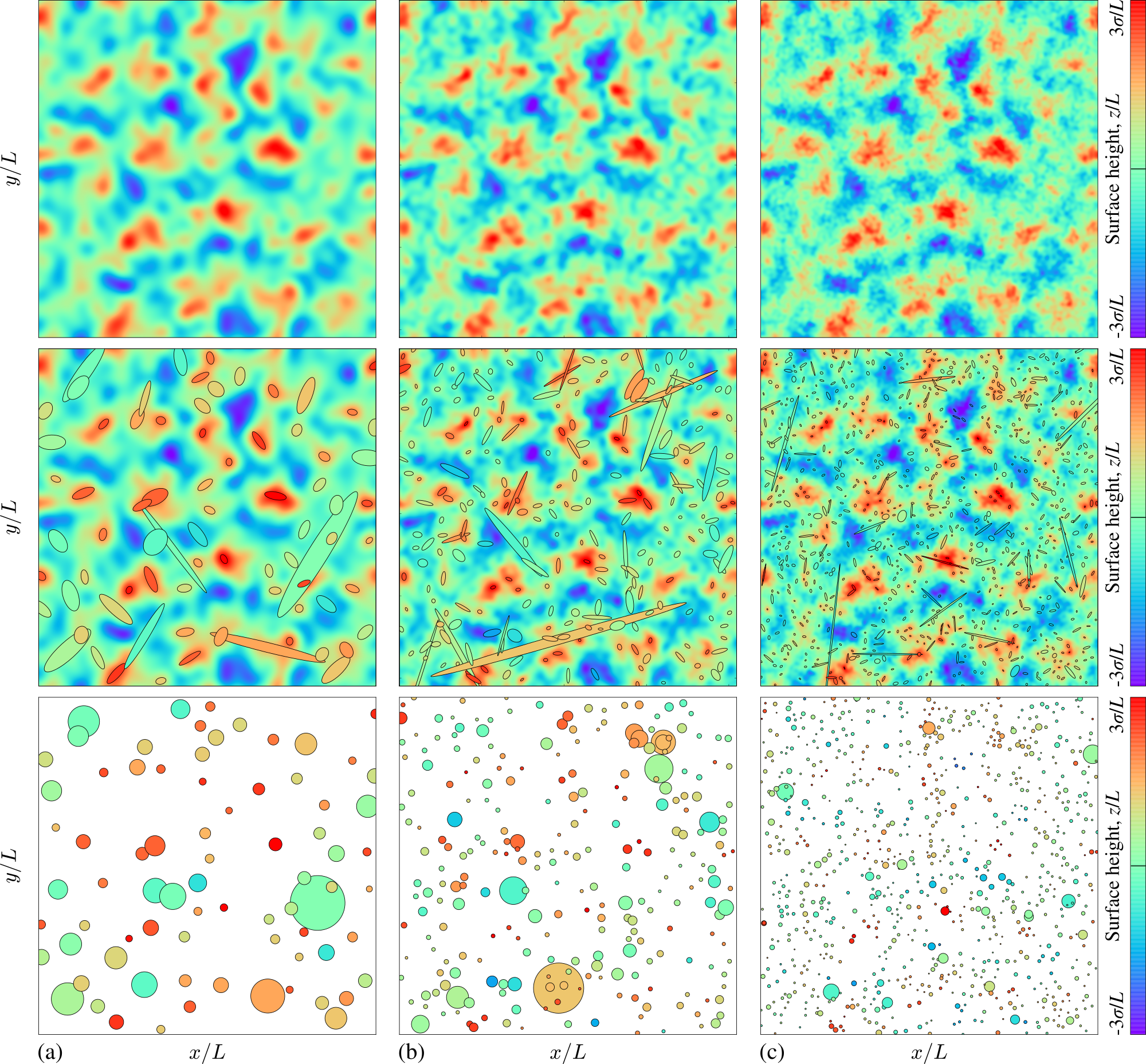}
 \caption{\label{fig:asperities}Generated rough surfaces (upper panel), detected elliptic asperities (middle panel) and approximate circular asperities with geometric mean curvature. For all surfaces, the Hurst exponent $H=0.8$, $L/\Delta x=2048$, lower cut-off $L/\lambda_l=4$, and root mean squared height $\sigma/L=0.025$: (a) $L/\lambda_s=16$, (b) $L/\lambda_s=32$, (c) $L/\lambda_s=64$. The surfaces are generated from the same white noise, therefore they are visually similar. The colour of asperities is selected according to the height of the summit.}
\end{figure}

In Fig.~\ref{fig:asperities} we show generated rough surfaces with different spectral content and the detected asperities approximated by ellipses with relevant orientation and circles with the geometric mean curvature.
Of course, the ellipses and circles give only an approximate representation of the roughness: anyway, the Hertzian theory is valid for arbitrary elliptic paraboloids, which can be approximated by ellipsoids near their summits.
The size of ellipses and circles corresponds to $L=1$ (l.u.\footnote{l.u. is length units.}) and root mean squared height (RMS height) $\sqrt{\langle (z-\bar z)^2\rangle}/L = \sigma/L =  0.025$. Obviously, a decrease of the RMS height would result in an increase of asperities' radii. It was shown by Greenwood~\cite{greenwood2006w}, that according to Nayak's theory~\cite{nayak1971tasme}, the asperities are only ``mildly'' elliptic, and thus, 
the geometric mean curvature can be successfully used in multi-asperity contact models~\cite{greenwood1985elliptic} instead of treatment elliptic contacts~\cite{bush1975w,johnson1985b}, without any significant loss in accuracy.
However, as seen by ``naked eye'' from Fig.~\ref{fig:asperities}, a considerable population of asperities has a pronounced ellipticity. 
Nevertheless, the mechanical analysis can be successfully carried out using simple geometric mean curvature and associated correction factors even for strongly elliptic asperities, see Table 2 in~\cite{greenwood2006w}.

\begin{figure}
  \includegraphics[width=1\textwidth]{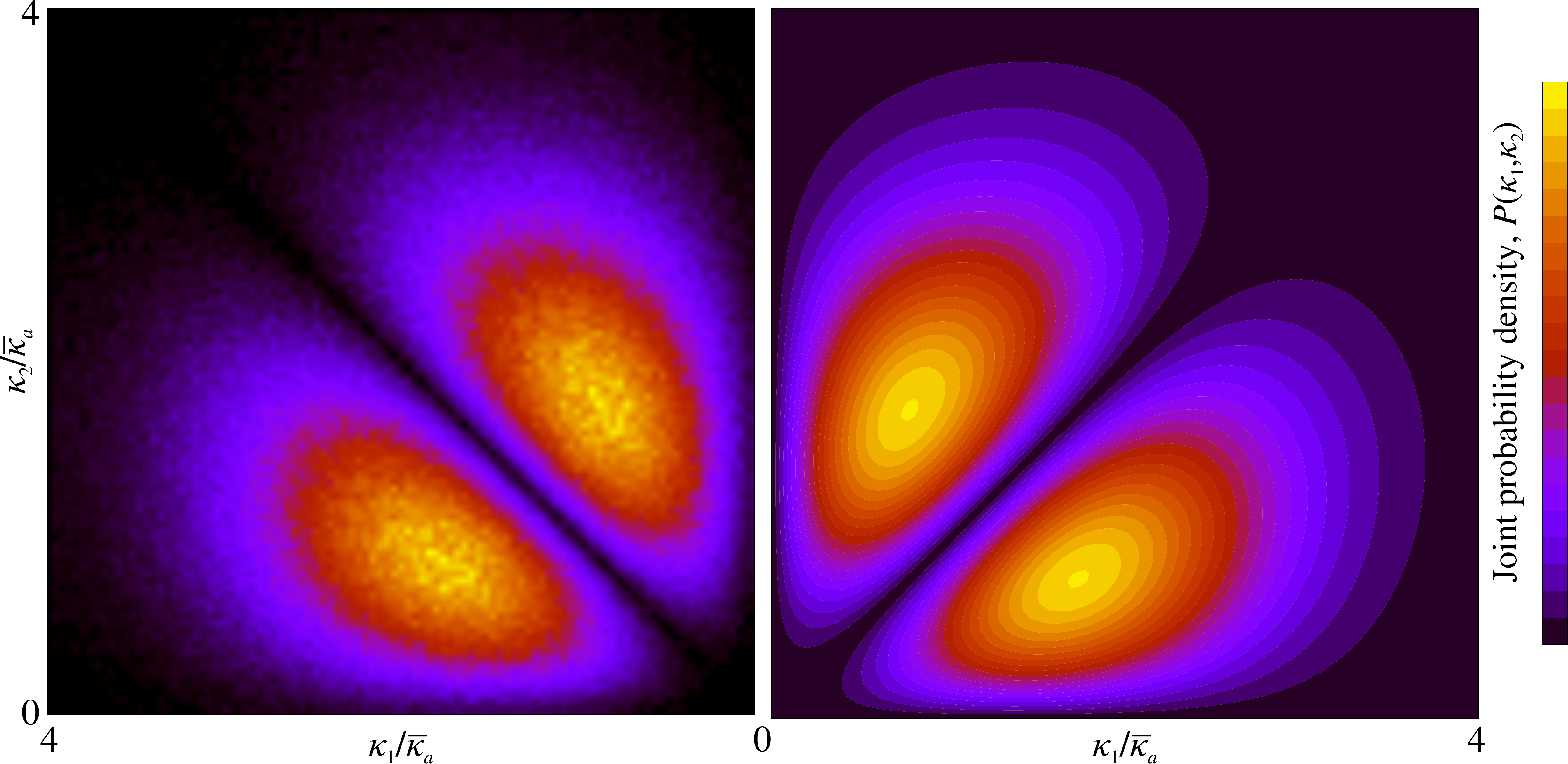}
 \caption{\label{fig:asp_stat}Comparison of the joint probability density computed numerically (on the left) and analytically, following Greenwood's equation~\eqref{eq:Greenwood}; principal curvatures $\kappa_1,\kappa_2$ are normalized by the average algebraic curvature $\bar\kappa_a$.}
\end{figure}

It seems that a good test for the surface generator would be a verification of agreement between asperities' statistics detected on the surface and Greenwood's analysis~\cite{greenwood2006w}, who obtained the following equation for the joint probability density of the ensemble of asperities' principal curvatures\footnote{In the original paper~\cite{greenwood2006w}, the probability density should be divided by a factor of two to fulfil the normalization.}:
\begin{equation}
  P(\kappa'_1,\kappa'_2) = \frac{27}{16\sqrt\pi}\kappa_1'\kappa_2'|\kappa_1'-\kappa_2'|\exp\left[-\frac{3}{16}\left(3\kappa_1^{\prime 2} + 3\kappa_2^{\prime 2} - 2\kappa'_1\kappa'_2\right)\right],
  \label{eq:Greenwood}
\end{equation}
where $\kappa_i' = \kappa_i/\sqrt{m_4}$ represent normalized curvatures.
This comparison is shown in Fig.~\ref{fig:asp_stat}. Numerically, the statistics is performed over an ensemble of asperities detected on a single realization of a surface with a rich spectrum $L/\lambda_l = 4$, $L/\lambda_s=2048$, $L/\Delta x = 8192$ and $H=0.8$; in total this surface contains $\approx 580\,000$ asperities. In accordance with Greenwood's findings, the most probable asperities possess the following ratio of principal curvatures $\kappa_1/\kappa_2 \approx 2.214$; the probability to find an asperity close-to-spherical is marginal. The overall mean geometrical curvature for the ensemble of asperities can be obtained from~\eqref{eq:Greenwood} by integration
\begin{equation}
  \bar\kappa'_g = \int\limits_0^\infty\int\limits_0^\infty \sqrt{\kappa_1'\kappa_2'} P(\kappa'_1,\kappa'_2) \,d\kappa_1'd\kappa_2' \approx 1.356
  \label{eq:mean_kg}
\end{equation}

\section{\label{sec:2}Numerical model}

The numerical model is based on a deterministic multi-asperity Hertzian contact with included long range elastic interaction between asperities. A set of asperities $\mathcal I$ with in-plane coordinates $x_i,y_i$, summit height $z_i$ and curvature radius $r_i$ is distributed on the top of an elastic half-space. The asperity parameters come directly from the generated rough surfaces. This set of elastic asperities comes in contact with a rigid sphere of radius $R$. The sphere's tip is located at $\{X_0,Y_0,Z_0\}$. 
Then the penetration between the sphere and $i$-th asperity is determined by:
\begin{equation}
  g_i = z_i  - \Delta z_i +\sqrt{R^2-(X_0-x_i)^2-(Y_0 -y_i)^2} - R - Z_0 = g_{i0} - \Delta z_i ,
  \label{eq:pen}
\end{equation}
where $\Delta z_i$ is the vertical displacement of the asperity induced by contact forces acting on other asperities of the surface, and 
$g_{i0} = z_i  +\sqrt{R^2-(X_0-x_i)^2-(Y_0 -y_i)^2} - R - Z_0$.
If the penetration is positive, Hertzian contact theory can be used to determine the resulting force $f_i$ and contact radius $a_i$ as:
\begin{equation}
  f_i = c_i\langle g_i\rangle^{3/2}, \quad a_i = \sqrt{r_i  \langle g_i\rangle}, \quad\mbox{ with } \quad c_i = \frac43 E^* \sqrt{r_i} %\sqrt{\frac{16}{9}  r_i E^{*2}}
  \label{eq:force}
\end{equation}
% {\color{red} CONSIDER SMOOTHING THIS FUNCTION FOR SHORT DISTANCES IN ORDER TO MAKE IT CLOSER TO VERTICAL DISPLACEMENT FIELD OF THE HERTZIAN SOLUTION}
where $\langle g_i\rangle = \max\{0,g_i\}$ denotes Macaulay brackets, therefore the forces are induced only if the penetration is positive.
We assume that the half-space's surface deforms under action of contact forces following Boussinesq solution, thus the vertical displacement due to a single concentrated force $f$ located at $x_f,y_f$ is given by:
\begin{equation}
  \Delta z_f = \frac{f}{\pi E^*\sqrt{(x-x_f)^2+(y-y_f)^2}}.
  \label{eq:dz}
\end{equation}
This coupling between additional displacement and forces makes the system of equations~\eqref{eq:pen}-\eqref{eq:dz} strongly coupled.
Then, following the superposition principle, at a given asperity $i$, its total vertical shift is given by:
$$
\Delta z_i = \frac{1}{\pi E^*}\sum\limits_{j \in \mathcal I\setminus i} \frac{f_j}{d_{ij}} =  \vec B \vec F,
$$
where $d_{ij} = \sqrt{(x_i-x_j)^2+(y_i-y_j)^2}$ and the symmetric interaction matrix $\vec B$ (containing the inverse distances) and the vector of forces $\vec F$ are given by:
$$
  \vec B = 
  \frac{1}{\pi E^*}\left[\begin{array}{ccccc}\displaystyle %a & a & a & a
       0 & d^{-1}_{12} & d_{13}^{-1} & \quad\dots & d_{1N}^{-1}\\[3pt]
        d^{-1}_{12} & 0 & d_{23}^{-1} & \quad\dots & d_{2N}^{-1}\\[3pt]
        d^{-1}_{13} &  d_{23}^{-1} & 0 & \quad\dots & d_{2N}^{-1}\\[3pt]
        \vdots & \vdots  & \vdots  & \ddots & \vdots \\[3pt]
        d^{-1}_{1N} \quad &  d_{N2}^{-1}\quad & d_{N3}^{-1} \quad& \quad\dots & 0\\[3pt]
           \end{array}
           \right], \quad \vec F = \left[\begin{array}{c}
                                          f_1\\[3pt]f_2\\[3pt]\\[3pt]\vdots\\[3pt] f_N
                                         \end{array}\right]
$$
Therefore from~\eqref{eq:force} we obtain a non-linear system of equations:
\begin{equation}
 \vec F - \vec C \langle \vec G_0 - \vec B \vec F\rangle^{3/2} = 0,
 \label{eq:sys}
\end{equation}
where $\vec C = \mathrm{diag}[c_1 \; c_2 \; \dots \; c_N]$ is the diagonal matrix, and the vector $\vec G_0 = [g_{10} \; g_{20} \; \dots \; g_{N0}]^\intercal$ contains the local penetrations in absence of elastic interactions. In used notations we also assume that Macaulay brackets applied to a vector and raising to $n$-th power imply the following operation with vector components:
$$
  \langle \vec V \rangle = [\langle \vec v_1 \rangle\; \langle \vec v_2 \rangle\; \dots \; \langle \vec v_N \rangle\;]^\intercal,\quad \vec V^n = [v_1^n \; v_2^n \; \dots \; v_N^n]^\intercal.
$$
Iterative explicit resolution of this problem (i.e. when to compute the configuration on the current load step, forces from the previous load steps are used to estimate the long-range interaction) in most cases leads to big errors.
On the other hand, in implicit integration, the system converges badly if Newton-Raphson method is used, because of the discontinuous nature of the function $\langle x\rangle$ and non-coercive properties. Therefore we regularize it as follows
$$
  m(x,\delta) =
  \left\{ \begin{array}{ll}
	      \displaystyle\frac{\delta^2}{2\delta-x} ,&\mbox{ if } x < \delta\\
	      x,&\mbox{ otherwise},
            \end{array}\right.
$$
If $\delta \to 0$ we recover the classical Macaulay brackets: $m(x,\delta) \to \langle x\rangle$. This approximation creates small repulsive forces between the sphere and asperities, but they decay quite rapidly (as $z^{-3/2}$), the parameter $\delta$ reduces to zero during the convergence loop enabling to ensure a better convergence for Newton-Raphson method, initial value of $\delta$ is chosen to be $\sigma/100$.
With this smoothing the system of equations~\eqref{eq:sys} reduces to
\begin{equation}
 \vec F - \vec C m( \vec G_0 - \vec B \vec F,\delta)^{3/2} = 0,
 \label{eq:sys_approx}
\end{equation}
whose solution tends to the solution of~\eqref{eq:sys} when $\delta\to0$. 
The initial guess for the force vector $\vec F_0$ is computed from a model without interaction ($\vec B = 0$), then from Eq.~\eqref{eq:sys}:
$$
  \vec F_0 = \vec C \langle \vec G_0 \rangle^{3/2}.
$$
A classical choice to take the initial guess from previous converged load step is not used since the possible set of active asperities changes from step to step (see next Section) and even if not, such initial guess often leads to divergence of the computational step.
From the initial guess, we obtain the increment of $\Delta \vec F$
as:
\begin{equation}
  \Delta\vec F = -\vec K^{-1}\vec R,
\end{equation}
where the free term $\vec R$ and the tangent matrix $\vec K$ are given by:
$$
\vec R = \vec F_0 - \vec C m( \vec G_0 - \vec B \vec F_0,\delta)^{3/2},
$$
$$
\vec K = \vec I - \frac32 \vec C m(\vec G_0 - \vec B \vec F_0,\delta)^{\frac12}\left.\ddp{m(\vec G_0 - \vec B \vec F,\delta)}{\vec F}\right|_{\vec F = \vec F_0},
$$
where $\vec I$ is identity matrix.
After every iteration, according to the classical Newton's method, the vector of forces is updated as $\vec F^i = \vec F^{i-1}+\Delta\vec F$ and all negative forces are removed as $\vec F^i = \langle \vec F^i\rangle$.
Regardless the fact that function $m(x,\delta)$ is smooth and well behaved, the convergence of the Newton method is still slow and non-monotonous, but it is rather robust.
The convergence is assumed to be reached when 
$$
  \|\vec \Delta \vec F^i \| / \|\vec F^i \| \le \varepsilon,
$$
the tolerance in all simulations is chosen to be $\varepsilon = 0.1$\%.

\subsection{Computational set-up}

We assume that the spherical indenter of radius $R$ is rigid and that its tip is centred at $x=X_0$, $y=Y_0$ of in-plane coordinates.
The substrate and asperities are made of gold with the following average elastic properties: the Poisson's ratio $\nu=0.42$ and the Young's modulus is $E=78$ GPa, giving the effective modulus $E^* \approx 94.71$ GPa. The size of the square region on elastic half-space with identified asperities is set to be $1$ mm $\times$ $1$ mm, i.e. the maximal contact radius is limited to $0.5$ mm.
The indentation simulations are carried out over a set of deformable asperities located on a deformable half-space; 
very small indentation depths are used. Naturally, no periodic boundary conditions are used.
The sphere is approached in 100 load steps from the separation\footnote{By separation we understand the distance between the sphere's tip and the mean line of the rough surface.}
 $u_z = 4\sigma$ down to $u_z = \sigma$ in 100 load steps. 
In order to accelerate the simulations, only those asperities are considered which 
(1) ``penetrate'' into the sphere in the undeformed state, i.e. $z_i > u_z + R-\sqrt{R^2 - (X_0-x_i)^2 - (Y_0-y_i)^2}$, and
(2) which are located in the circle $\sqrt{(x-X_0)^2 + (y-Y_0)^2} < r_0$, where $r_0 = 1.5\sqrt{2(4\sigma - u_z)R}$, which is 1.5 times bigger than the radius of geometrical overlap of a sphere, whose tip is  penetrated under a plane by distance $(4\sigma-u_z)$.

After convergence, at every step we record the total contact force $N$ and total contact area $A$. 
Moreover, at every load step, a coarse grained map of the contact pressure and the contact area is saved. 
The map represents a regular grid $N\times N$ created over the indentation area $L\times L$ with cells of size $\Delta x \times \Delta x$. 
In each cell $\{i,j\}$, $i,j \in [0,N-1]$ all contacting asperities 
$\mathcal I_{ij} \subset \mathcal I$ 
contribute to the coarse-grained contact area fraction $A_{ij}$, and to the coarse-grained contact pressure $p_{ij}$, where 
\begin{equation}
  A'_{ij} = \frac{1}{\Delta x^2}\sum\limits_{k\in\mathcal I_{ij}} \pi a_k^2,\quad p_{ij} = \frac{1}{\Delta x^2} \sum \limits_{k\in\mathcal I_{ij}} F_{k},
  \label{eq:coarse}
\end{equation}
where $a_k$ is computed using~\eqref{eq:force} and $\mathcal I_{ij}$ is a set of asperities with in-plane coordinates located in the given cell.

\section{\label{sec:3}Results}

\subsection{Two regimes of the contact-area growth}

From the physical evidence, under most conditions and for most materials, the frictional force (static friction) increases practically linearly with the normal load independently on the nominal contact area, which is reflected by the Amontons-Coulomb's friction law. Accordingly to the adhesive theory of contact~\cite{bowden_tabor}, the frictional force is proportional to the true contact area. Therefore, the linear growth of the contact area could be considered as a necessary condition of the model validity. The model of Greenwood and Tripp~\cite{greenwood1967elastic}, which represents the fully analytical treatment of the elastic indentation problem of a rough surface including elastic interaction between contacting asperities in the statistical sense , predicts the linearity between the force and the contact area, as well as the recent full scale numerical treatment of this problem, which was carried out by Pastewka \& Robbins~\cite{pastewka2016contact}. However, as remarked in~\cite{greenwood1970contact,pastewka2016contact}, at very small loads/large separations a single-asperity contact regime exists, which results in  Hertzian scaling for elastic asperities\footnote{If one assumes that the elastic asperities readily plastifies, the linearity can be preserved even at the first contact, since the contact pressure readily saturates at the hardness level and thus the contact area grows proportionally to the contact force.}. 

\begin{figure}
 \centering\includegraphics[width=1\textwidth]{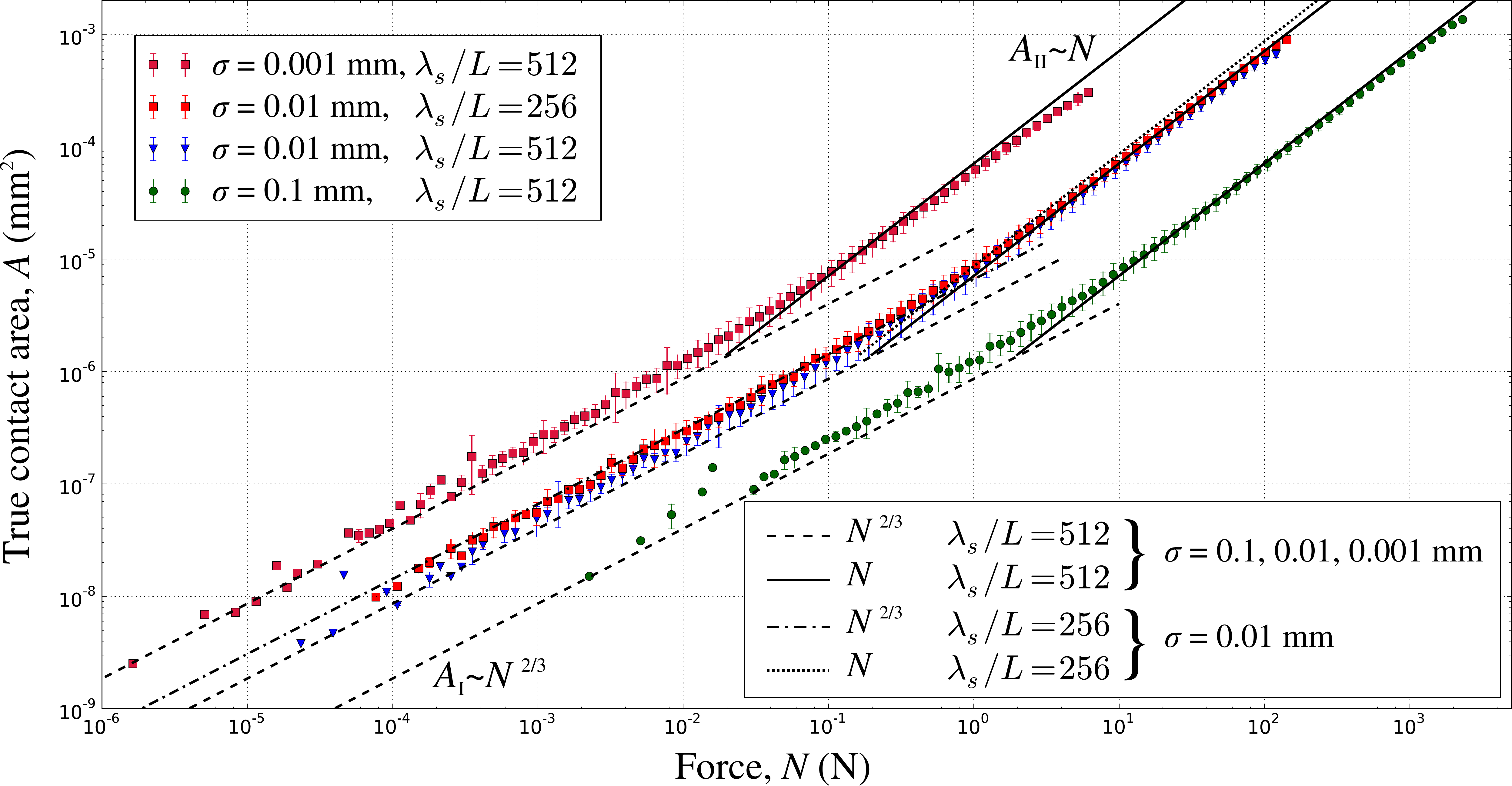}
 \caption{\label{fig:area}Evolution of the contact area for different root mean squared roughness $\sigma=\{0.001,0.01,0.1\}$ and different upper cut-offs $L/\lambda_s=\{256,512\}$; to guide the eye, the $2/3$-power law and a linear evolution with relevant factors (Eqs.~\eqref{eq:AI},\eqref{eq:AII}) are also plotted.}
\end{figure}

Expectedly, our numerical study with the deterministic multi-scale model  identifies these two deformation regimes (see Fig.~\ref{fig:area}). 
The data were averaged over 120 simulations per each combination of root mean squared roughness $\sigma=\{0.001,0.01,0.1\}$ and different upper cut-offs $L/\lambda_s=\{256,512\}$, the lower cut-off was kept constant $L/\lambda_l = 24$. 
The vertical displacement of the indenter's tip was changed from $Z_0 = 4\sigma$ to $\sigma$ in 100 load steps, the indenter's radius is fixed to be $R=2.5$ mm.
The error bars correspond to a single standard deviation; the fact that error bars are missing in very low pressure-regime is explained by the fact that only in a single realization the contact occurs at such small loads.

The first deformation regime corresponds to the deformation of a single Hertzian asperity, it occurs at very light loads and the contact-area grows as  $A_I \sim N^{2/3}$. The second regime is characterized by an approximately linear evolution of the contact area $A_{II}\sim N$ and it corresponds to multi-asperity statistical contact. The first regime then fully depends on the curvature of the first contacting  asperity, which can be assumed to be the average geometrical mean curvature over all asperities\footnote{A more accurate estimation only for highest asperities could be obtained using Eq. (11) from~\cite{greenwood2006w} and by correcting it with a factor $1/2$.} $r = \bar r_g \approx 0.737/\sqrt{ m_4}$ (the last approximate equality follows from~\eqref{eq:mean_kg}):
\begin{equation}
  A_I = \pi \left(\frac{3\bar r_g}{4E^*}\right)^{\!\!\frac 2 3} N^{^{\frac 2 3}} \approx \pi \left(\frac{0.55275}{\sqrt{ m_4}E^*}\right)^{\!\!\frac 2 3} N^{^{\frac 2 3}}.
\label{eq:AI}
\end{equation}
In the second regime, at moderate loads, the contact area should grow accordingly to classical multi-asperity theories~\cite{greenwood2006w,bush1975w,carbone2008jmps}, Persson's theory~\cite{persson2001jcp} and more recent full-scale numerical simulations~\cite{hyun2004pre,pei2005jmps,putignano2012jmps,afferrante2012w} with the scaling depending on the root mean squared roughness gradient $\langle |\nabla z|^2\rangle$, which for Gaussian isotropic surface can be estimated through the second spectral moments $\langle |\nabla z|^2\rangle = \sqrt{2m_2}$ (see~\ref{app}). Therefore, in the second regime the contact area growth is determined by the following equation:
\begin{equation}
 A_{II} = \frac{\beta\kappa}{\sqrt{2 m_2}E^*} N,
\label{eq:AII}
\end{equation}
where the constant $\beta \approx 0.2$ was identified from our simulations to fit approximately the data. The factor $\kappa=\sqrt{2\pi}$ appears from statistical multi-asperity models~\cite{bush1975w,greenwood2006w,carbone2008jmps} and was shown to be very accurate for predicting the contact area evolution at small loads for nominally flat surfaces~\cite{yastrebov2017jmps}. For higher loads, both multi-asperity models~\cite{carbone2008jmps,greenwood2006w} and numerical simulations~\cite{paggi2010w,yastrebov2017jmps} demonstrate a slightly non-linear area growth which depends on Nayak's parameter $\alpha = m_0m_4/m_2^2$: greater $\alpha$ results in smaller contact area.
 Note that in our simulations we remain far from the third deformation regime, purely Hertzian one, which was identified in~\cite{pastewka2016contact}.

\subsection{Transition between the regimes}

The transition force $N^*$ between the two regimes can be found by assuming equality of areas $A_I(N^*) = A_{II}(N^*)$, which results in the following expression:
\begin{equation}
  N^* = 0.55275^2 \frac{\pi^3 (2  m_2)^{3/2} E^*}{\beta^3\kappa^3  m_4} \approx 212.66  \frac{  m_2^{3/2} E^*}{ m_4},
  \label{eq:trans}
\end{equation}
where the numerical factor $212.66 \approx 0.55275^2 (\pi\sqrt2/\kappa \beta)^3$ with $\kappa = \sqrt{2\pi}$ and $\beta = 0.2$. Note that in~\cite{pastewka2016contact}, a different expression was obtained\footnote{Lower index ``PR'' was used to indicate the authors (Pastewka \& Robbins) of the mentioned paper~\cite{pastewka2016contact}, we show here the combination of Eq.(5) and (6) from this paper.}: $N^*_{\mathrm{PR}} = (9\pi^3/4m_4) (\sqrt{2 m_2}/\kappa)^3E^*$, which after evaluating all numerical quantities and assuming $\kappa = \sqrt{2\pi}$ gives $N^*_{\mathrm{PR}} \approx 12.5287 m_2^{3/2}  E^*/m_4$ which is considerably smaller than our value~\eqref{eq:trans}. This is because of slight overestimation made by the authors~\cite{pastewka2016contact} of the asperity radius $\bar r_g = 2/\sqrt{m_4}$ and of non-taking into account the scaling factor\footnote{This scaling factor is implicitly contained in authors'~\cite{pastewka2016contact} estimation of the mean contact pressure $N/A_{\mbox{\tiny Hertz}}$, which disappears in the transition to the fully Hertzian regime at high loads, but had to be preserved at light loads, which was not the case.} $\beta$ which was required in our case to predict the contact area growth in the second regime.
On the other hand, Pastewka \& Robbins' expression in its final form lacks the fourth spectral moment $m_4$, which in their case was expressed through the short wavelength cut-off, Hurst exponent and the second spectral moment. Returning to our results, we show that the lower cut-off $\lambda_l$ affects slightly the numerical values, but keeps the transition interval unchanged. The normalized data plotted in Fig.~\ref{fig:area_norm} include the results found for different indenter's radius $R=\{2.5;\;5\}$ mm, and different lower cut-off wavelength $\lambda_l/L = \{24;\;48\}$, which seemingly do not change the transition point when normalized.

\subsection{Scaling}

\begin{figure}
 \centering\includegraphics[width=1\textwidth]{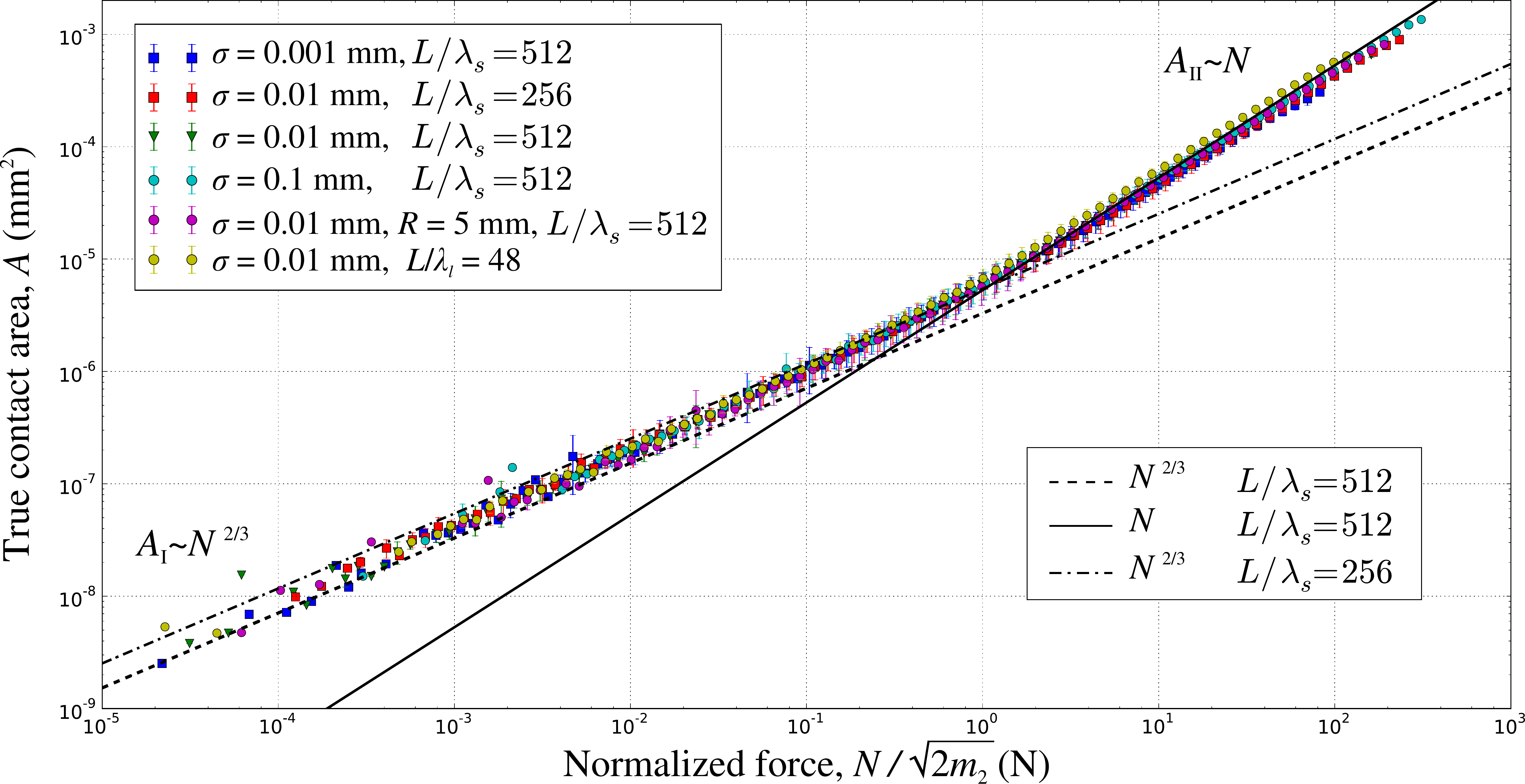}
 \caption{\label{fig:area_norm}Evolution of the contact area as a function of the force normalized by the RMS gradient, the data are obtained for different root mean squared roughness $\sigma=\{0.001,0.01,0.1\}$ and different upper cut-offs $L/\lambda_s=\{256,512\}$; if it is not stated differently $L/\lambda_l = 24$ and $R=2.5$ mm; to guide the eye, the $2/3$-power law and a linear evolution with relevant factors (Eqs.~\eqref{eq:AI},\eqref{eq:AII}) are also plotted.}
\end{figure}

Since all spectral moments scale proportionally to the roughness variance $m_4 \sim m_2 \sim m_0 = \sigma^2$ then the contact areas in the first~\eqref{eq:AI}  and the second~\eqref{eq:AII} regimes scale as follows:
\begin{equation}
  A_I(N,\sigma) = A_I(N/\sigma) \sim \left(\frac{N}{\sigma}\right)^{2/3},\quad A_{II}(N,\sigma) =  A_{II}(N/\sigma) \sim \frac N \sigma.
\end{equation}
The scaled results and theoretical predictions are plotted in Fig.~\ref{fig:area_norm}.
Note that the scaling by the RMS gradient is used which is equivalent. In the geometrical model, scaling the RMS roughness scales proportionally the RMS gradient and curvature, however, it should be bared in mind that in real life, changing RMS roughness (removing higher asperities), in general, does not result in changing the RMS gradient which depends on small scale roughness, therefore the latter was used to scale the results. According to this scaling, the transition load between the  two regimes should be proportional to the RMS height $N^* \sim \sigma$.

\subsection{Maps of contact area/pressure}

The linear regime~\eqref{eq:AII} seemingly follows directly from classical theories.
If one assumes that a smooth Hertzian solution can be used at macro scale, and that at the micro scale the contact area is proportional to the local contact pressure, then the total contact area is proportional to the normal force.
Let the local contact area $dA$ be given by $dA(x,y) = \kappa p(x,y)dA_0/(\sqrt{2m_2}E^*)$ and the pressure is distributed according to Hertz theory $p(x,y) = p_0\sqrt{1-(x^2+y^2)/a^2}$, with $p_0 = 3N/(2\pi a_0^2)$, where $N$ is the squeezing force. Since the integral of pressure over the contact area is simply the resulting force $N$, then the contact area is given by:
\begin{equation}
  A_{\mathrm{th}} = \frac{\kappa}{\sqrt{2m_2}E^*}N .
  \label{eq:Ath}
\end{equation}
However, the contact area for indentation of the rough contact~\cite{greenwood1967elastic} can be significantly larger than the one predicted by Hertzian theory in case of light indentations, therefore the separation between the micro and macro scales cannot be assumed.
The apparent contact area is thus not easy to determine. In theory, because of the infinite support of the Gaussian distribution, the contact has a non-zero probability to occur at any distance $r < R$ from the contact centre.
In case of deterministic multi-asperity model, the contact spots appear at random locations outside the Hertzian contact zone. Since the apparent contact area $A_0$ is unknown, the mean contact pressure $\bar p = N/A_0$ is also unknown as well as the contact area fraction $A/A_0$. However, the latter can be evaluated locally with our numerical model by using the stored coarse grained data. 
The striking difference between the theoretical prediction~\eqref{eq:Ath} and the numerical results~\eqref{eq:AII}, comes from the fact that the former assumes non-interacting asperities and also assumed the scale separation and the validity of Hertz's theory at macro-scale, which visibly does not hold. 

\begin{figure}
 \centering\includegraphics[width=1\textwidth]{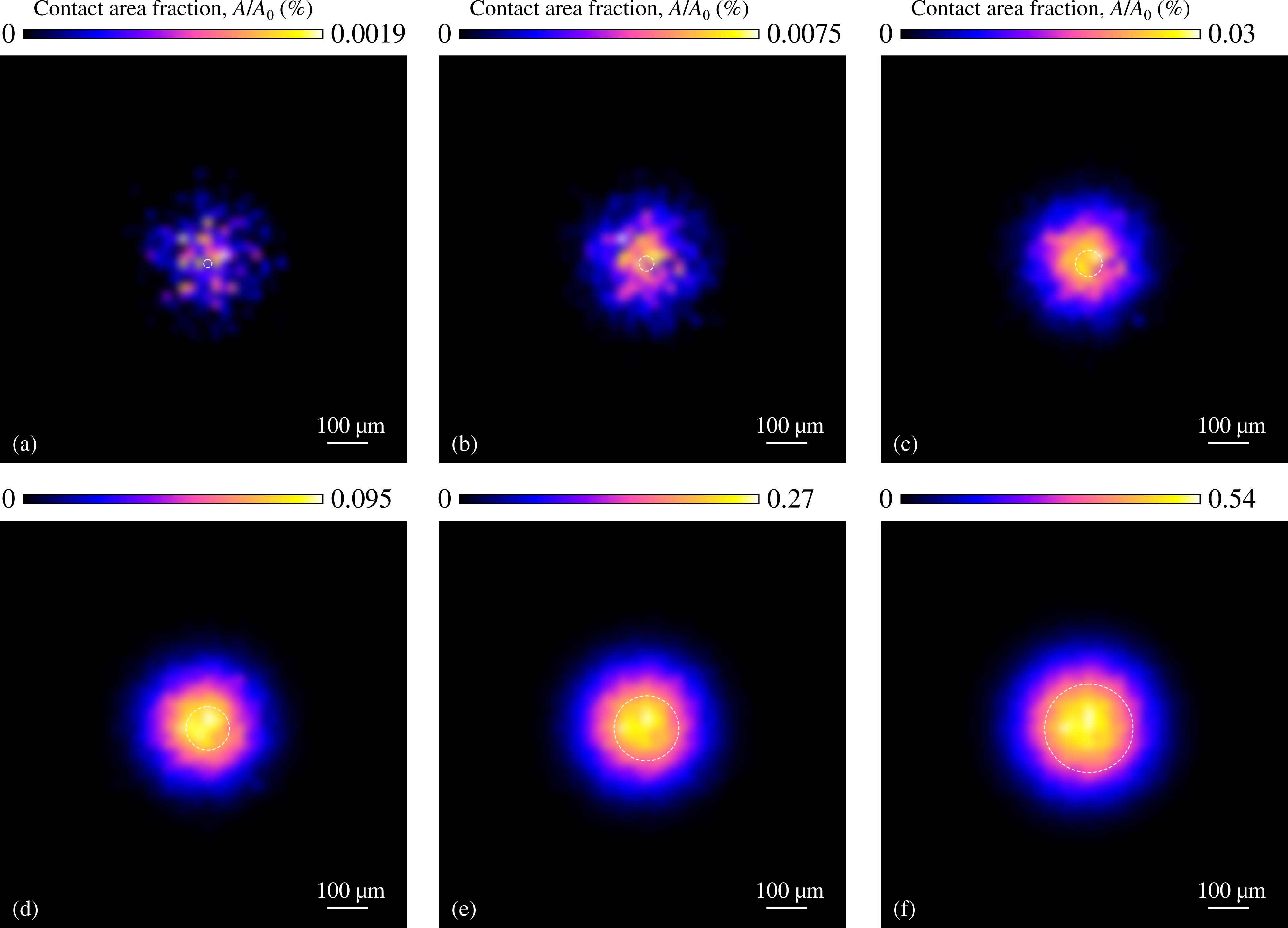}
 \caption{\label{fig:map_area}Evolution of the contact area fraction under the spherical indentation: the brighter colours correspond to bigger contact area fraction: sub-figures (a-f) correspond  to sphere's tip position 
 $Z_0/\sigma \approx \{4,\;3.57,\;3.14,\;2.71,\;2.29,\;1.86\}$, respectively. The black square is of size $1\times1$ mm. The white dashed line denotes the outline of the Hertzian contact area's prediction for the mean corresponding contact force $N$.}
\end{figure}

The coarse grained and smoothed with bi-cubic B\'ezier interpolation results for the spatial area distribution are presented in Fig.~\ref{fig:map_area} and compared with Hertzian contact radius for the same force. 
These results are obtained by averaging the coarse grained results over 2000 simulations with $R=2.5$ mm, $\sigma=10$ $\mu$m, $L/\lambda_l=512$, $L/\lambda_s=24$, $H=0.8$ at different indentations depth ranging from $Z_0/\sigma=4$ to $Z_0/\sigma\approx1.4$, the latter corresponds to the contact force $N\approx116.6$ (N).
At very light loads the expansion of the contact zone is remarkably bigger than the Hertzian prediction. This discrepancy decreases under increasing load, but still remains big in the studied loading interval. 

\begin{figure}
 \begin{center}
   \includegraphics[width=1\textwidth]{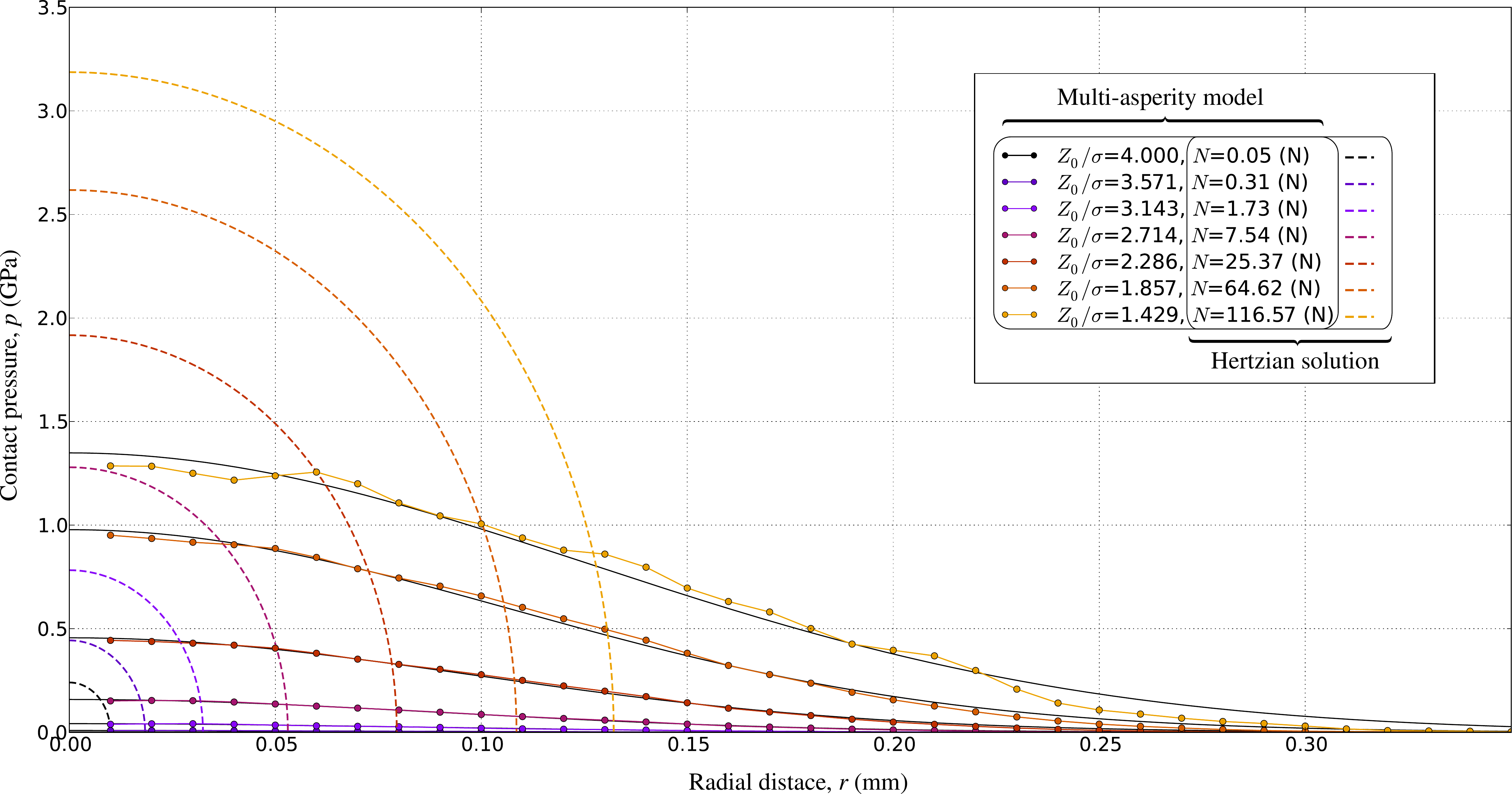}\\\footnotesize(a)\\[5pt]
   \includegraphics[width=1\textwidth]{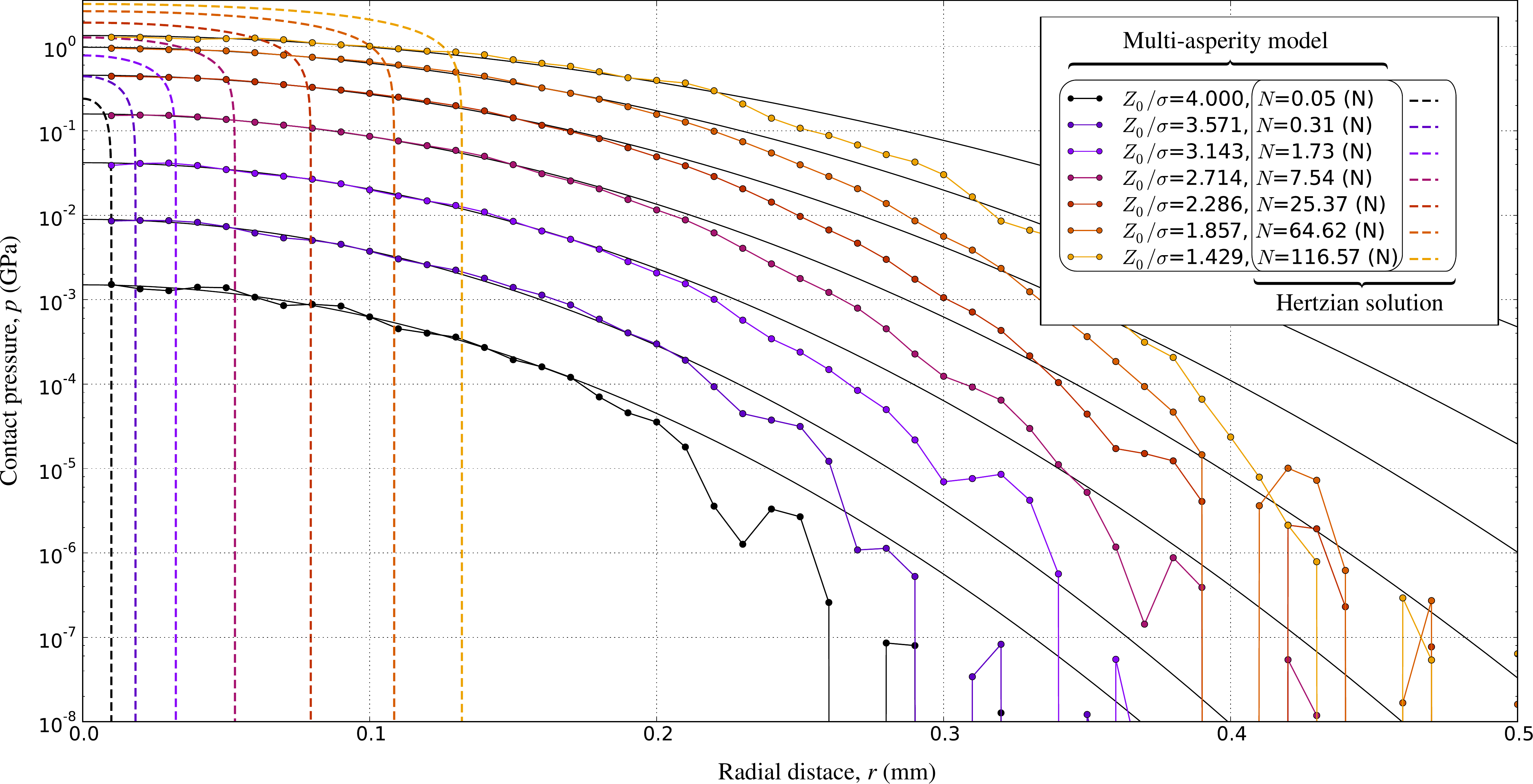}\\(b)\normalsize
  \caption{\label{fig:radial_pres}Contact pressure distribution in rough contact for different loads compared with Hertzian solution (dashed curves of the same colour) for the same contact force (a) normal scale, (b) semi-log scale. Black solid lines correspond to Gaussian fit.}
 \end{center}
\end{figure}

To compute the radial distribution of the contact area fraction and contact pressure at different loads, we use the coarse grained data~\eqref{eq:coarse}: $A'_{ij} = A'(x_i,y_j)$, $p_{ij} = p(x_i,y_j)$, where $i,j \in [0,N-1]$ and $x_i = (i+0.5)dx$, $y_j = (j+0.5)dx$. In our simulations, the following coarse-graining was used $N=50$ and $dx=20$ $\mu$m. To find the average radial data, we first interpolated $A'_{ij}$ and $p_{ij}$ in between grid points using bi-cubic B\'ezier interpolation to obtain $\tilde A'(x,y)$, $\tilde p(x,y)$, then we averaged over concentric circles centred at the indenter's tip $X_0,Y_0$ to obtain mean radial values $A(r),p'(r)$, which assumes that in statistical sense the both quantities should be axisymmetric:
$$
 A'(r) = \frac{1}{2\pi}\int\limits_0^{2\pi} \tilde A'(X_0+r\cos(\phi),Y_0+r\sin(\phi)) \,d\phi,
$$
$$
 p(r) = \frac{1}{2\pi}\int\limits_0^{2\pi} \tilde p(X_0+r\cos(\phi),Y_0+r\sin(\phi)) \,d\phi.
$$
The resulting radial pressure distributions at different load levels are presented in Fig.~\ref{fig:radial_pres}(a,b) in normal and semi-log scales, respectively.  An expected resemblance of our results with those presented in Fig.5(a) from Greenwood and Tripp's analysis~\cite{greenwood1967elastic} can be noticed.
The pressure distribution resembles Gaussian one, but it decays faster than the Gaussian after approximately two times standard deviation from the centre. The extension of the contact zone beyond the Hertzian contact is even more evident in the figures of radial distributions. The contact area fraction which follows approximately the same decay is not shown. It could be noticed that for selected contact characteristics the contact area fraction does not go beyond 1\%  even at the highest considered load $N\approx116.6$ (N). Therefore the use of the asperity based model can be justified.

\subsection{Probability distributions}

\begin{figure}
 \begin{center}
   \includegraphics[width=1\textwidth]{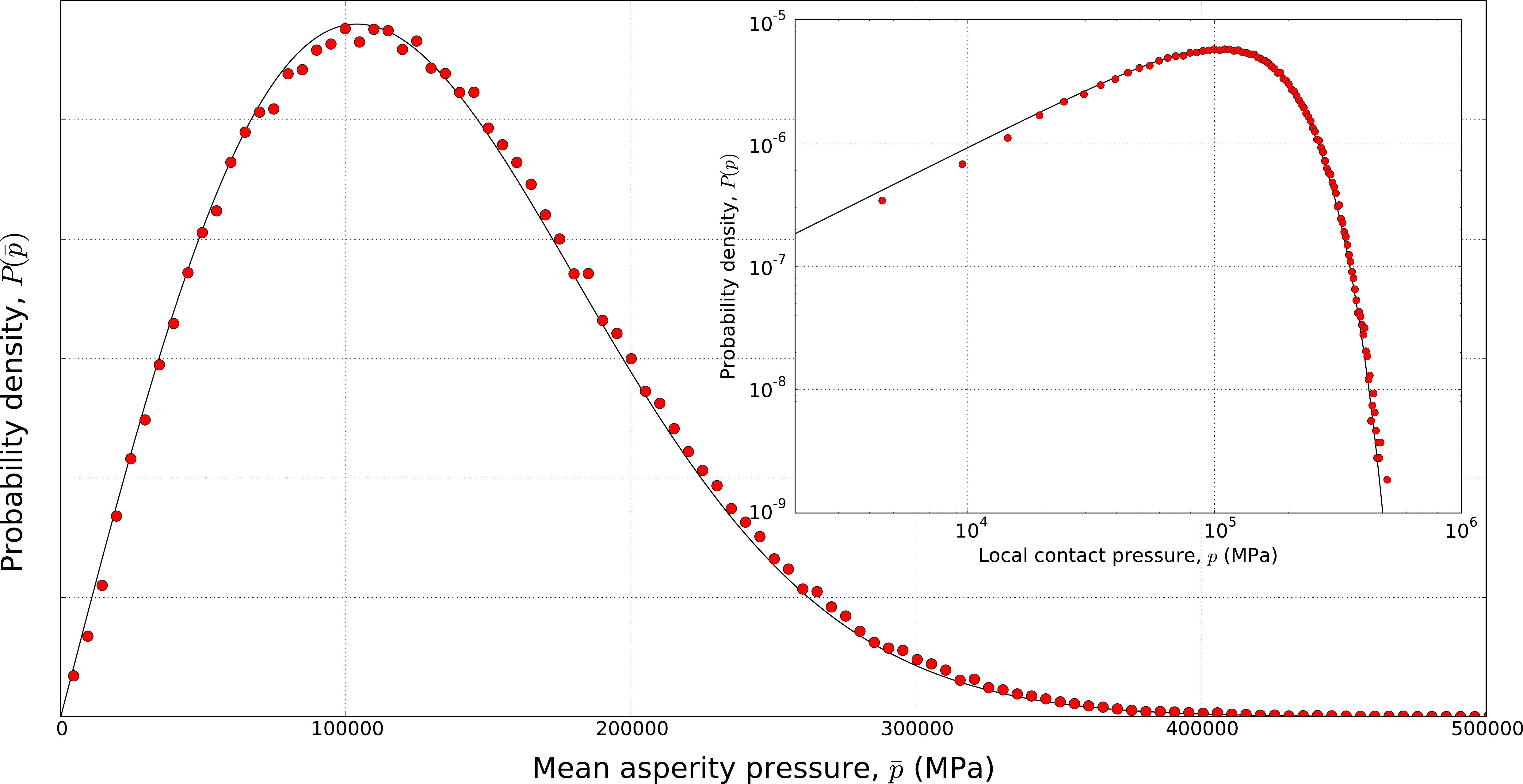}
  \caption{\label{fig:pres_probability}Mean asperity pressure probability $\bar p$ evaluated over 2000 simulations for $Z_0/\sigma = 2$ shown in normal and log-scale (see inset). Black lines correspond to fitted Rayleigh distribution~\eqref{eq:R}.}
 \end{center}
\end{figure}

The deformation regime of individual asperities can be determined by the mean contact pressure at every asperity $\bar p_k = F_k/(\pi a_k^2)$. Therefore it could be useful to construct the probability density function of mean contact pressures $P(\bar p)$, which are experienced at individual asperities. To do so, we carried out simulations for different indentation depths $Z_0/\sigma = \{3,\, 2.5,\, 2\}$ with $R=2.5$ mm, $\sigma=10$ $\mu$m, $L/\lambda_l=512$, $L/\lambda_s=24$, $H=0.8$;
for every distribution, the data of 2000 simulations were used, which results in $\approx 320\,000$, $157\,000$, $39\,000$ data points for $Z_0/\sigma = \{2,\, 2.5,\, 3\}$ indentation depths, respectively. 
The results for $Z_0/\sigma = 2$ are shown in Fig.~\ref{fig:pres_probability}, the data look the same for other indentation depths.
The numerical data fit very accurately the Rayleigh distribution: 
\begin{equation} 
P_R(\bar p) = \frac{\bar p}{\Delta^2}\exp\left(-\frac{\bar p^2}{2\Delta^2}\right),
\label{eq:R}
\end{equation}
where $\Delta$ is the scaling parameter of pressure units.
Clearly, the pressures at asperities are too huge to retain the deformation within the elastic regime. Normally, the maximal contact pressure in the interface is determined by material hardness $H$, which for elasto-plastic materials can be approximately assumed to be $H\approx3\sigma_Y$, where $\sigma_Y$ is the plastic yield stress~\cite{tabor2000hardness,johnson1985b,hill1989theoretical,mesarovic1999spherical,kogut2002elastic}. The yield stress for gold ranges between $100$-$200$ MPa and the Brinnel hardness ranges in $190-250$ MPa. However, at the scale of asperities the indentation size effect can be pronounced. For spherical indentation, it can be taken into account in terms of hardness change with the indenter's radius $R$~\cite{swadener2002correlation}:
$$
 H = H_0\sqrt{1+\frac{R^*}{R}},
$$
where $R^* = \bar r/(\rho_s b)$, and $\bar r$ is the Nye factor\footnote{The Nye factor is a dimensionless factor determining how much greater dislocations should be created to accommodate plastic deformation compared to the number of geometrically necessary dislocations}, $\rho_s$ is the density of statistically stored dislocations, and $b$ is the Burger's vector. The Burger's vector in FCC gold with lattice constant $a=0.408$ nm, is given by $b=a/\sqrt2 \approx 0.288$ nm, the density of statistically stored dislocations in strongly deformed region under the indentation could be estimated to be $\rho_s \approx 10^{16}$ m$^{-2}$, taking $\bar r = 2$, we obtain $R^* \approx 0.69$ $\mu$m, which is in good agreement with the data presented in~\cite{kim2018indentation}. Assuming that the mean asperity radius is $\bar r_g \approx 0.737/\sqrt{\bar m_4}$ and assuming macroscopic hardness $H_0=250$ MPa, we obtain the micro-asperity hardness to be
$$
  H = 250 \sqrt{1 + 0.69/\bar r_g}  %\text{(MPa)},
$$
with $\bar r_g$ measured in micrometers. For the given example with $L/\lambda_l=24$, $L/\lambda_s=512$ and $H=0.8$ and $L=1$ mm, we obtain $\bar r_g \approx 0.18$ $\mu$m, raising the micro-hardness to $H\approx 1.2$ GPa, which is still very small compared to the pressure range going up to 0.5 TPa. Definitely, almost all contacting asperities would be deformed plastically, therefore including elasto-plastic asperity behaviour would be needed for a predictive modelling.

\begin{figure}
 \begin{center}
   \includegraphics[width=1\textwidth]{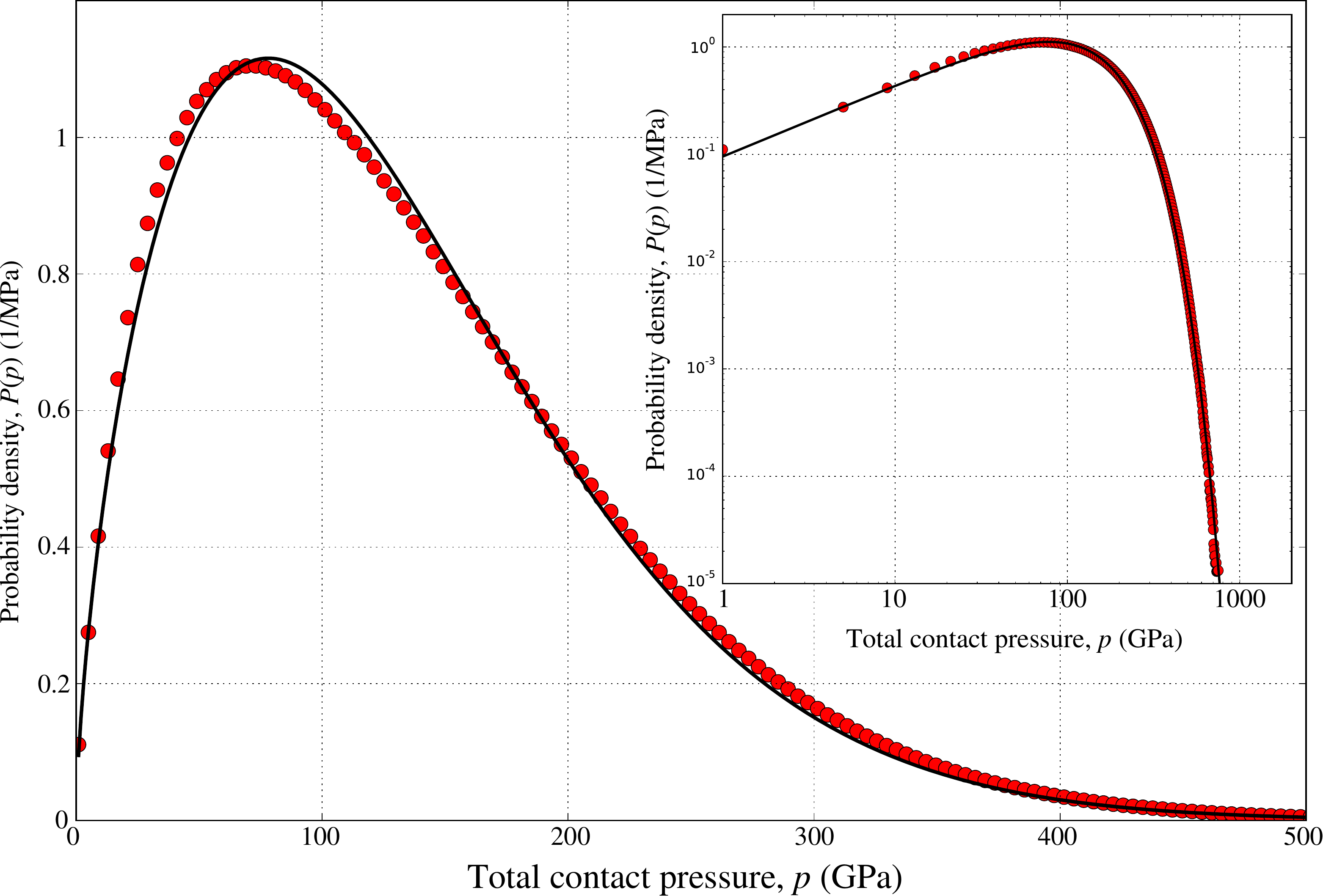}
  \caption{\label{fig:full_distrib}Probability density of the total pressure distribution evaluated over 2000 simulations at $Z_0/\sigma=2$. Black lines correspond to Eq.~\eqref{eq:my_distr} with $\alpha=2/3$ and $\beta=3/2$.}
 \end{center}
\end{figure}

The full pressure distribution can be obtained from this distribution. At every asperity, the Hertzian pressure distribution results in a linear probability density $P_a(p)$:
\begin{equation}
P_a(p) = 
  \left\{ \begin{array}{ll}
	      \displaystyle 2p/p_0^2 ,&\mbox{ if } p \le p_0\\
	      0,&\mbox{ otherwise},
            \end{array}\right.
            \label{eq:asp_pdf}
\end{equation}
where $p_0$ is the maximal pressure at this asperity, which in turn is given by $p_0 = 1.5\bar p$. The resulting full pressure distribution is shown in Fig.~\ref{fig:full_distrib}. Evidently, in the limit of zero pressure the global contact pressure distribution should be linear as it will present a sum of linear contribution from few individual asperities~\eqref{eq:asp_pdf}. However, because of the abrupt decay of these probability densities at given $p=p_0$, the global probability density has a slower growth than linear one in the Rayleigh distribution, the decay is also slower than Gaussian. Our simulations suggest, that the global probability distribution is given by the following function:
\begin{equation}
P_{gl}(p) = \gamma p^\alpha \exp\left[-(p/\delta)^\beta\right],
\label{eq:my_distr}
\end{equation}
where we identified $\alpha\approx2/3$ and $\beta\approx3/2$. The fact that $\alpha \approx 2/3$ is consistent with recent numerical findings~\cite{wang2017gauging} where the author identified $\alpha\approx0.7$, which at the same  contradicts the Persson's model which predicts a linear slope for relatively ``small'' pressures. The sub-Gaussian decay for high pressures can probably be explained by the fact that we consider contact between a sphere and flat rough surface and not between nominally flat surfaces as it is usually the case for the rough contact analyses.

\begin{figure}
 \begin{center}
   \includegraphics[width=1\textwidth]{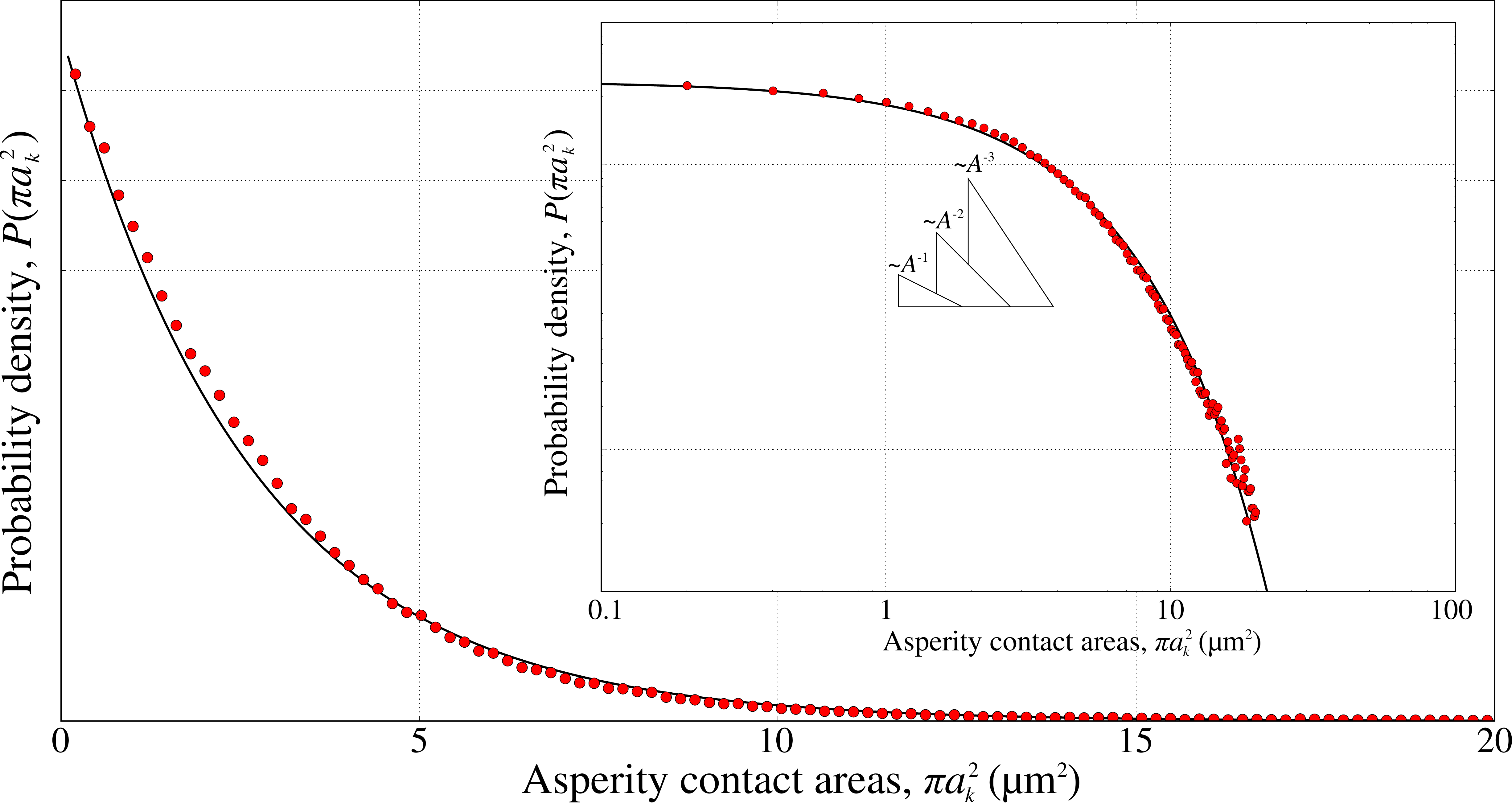}
  \caption{\label{fig:area_distr}Probability density of contact areas associated with individual asperities (normal scale and log-log in the inset), the distribution evaluated over 2000 simulations at $Z_0/\sigma=2$. Black lines correspond to exponential decay $P(A) \sim \exp(-A/a)$}
 \end{center}
\end{figure}

The distribution of individual contact spot sizes associated with contacting asperities is of big importance for adhesive wear~\cite{bowden_tabor,frerot2018mechanistic} as well as for electric and thermal conductivity~\cite{greenwood1966constriction,slade2017electrical}. The distribution of areas evaluated for same parameters and for indentation depth $Z_0/\sigma=2$ is plotted in Fig.~\ref{fig:area_distr} and follows exponential decay $P(A) \sim \exp(-A/a)$. 
Our results share the same features with experimental observations~\cite{dieterich1996imaging} and numerical results~\cite{hyun2007elastic,muser2017meeting,frerot2018mechanistic}, namely a ``plateau'' at small areas  in log-log scale and a faster decay at larger areas. However, we do not observe a power law decay in a considerable interval of areas, as was observed in all aforementioned studies. The decay is purely exponential. It can be justified by the fact that individual contact clusters associated with separate asperities never merge in our model, therefore the number of bigger clusters is lower that what is observed in the literature for stronger loads and areas, which considerably overpass  $\approx0.5$ \% of contact area fraction obtained at these loading conditions with our model.

% \section{Comparison}
% 
% Generate rough surfaces
% detect asperities and use deterministic asperity based models
% Do BEM on it 
% And use statistical models (BGT, simplified elliptic model, Persson)
% 
% Compare results
% 
% Pastewka L, Robbins MO (2016) Contact area of
% rough spheres: Large scale simulations and simple
% scaling laws. Appl Phys Lett 108(221601):1-5
% 
% + 
% 
% Borri-Brunetto, Ciavarella ``Elastic indentation of a rough surface by a conical punch''

\section{\label{sec:4}Discussion}

The deterministic multi-asperity model inspired from~\cite{afferrante2012interacting} enables to solve precisely the mechanical normal contact of two spheres in presence of roughness. The semi-analytical solution~\cite{greenwood1967elastic} of  this problem was derived by Greenwood and Tripp in late 60's right after the pioneering work of Greenwood and Williamson on mechanical contact of rough elastic half-spaces~\cite{greenwood1966prcl}. In their solution, the elastic interaction between asperities is taken into account in a statistical sense via an iterative numerical scheme. Recently, Pastewka and Robbins revisited this problem using deterministic full-scale simulations of elastic contact between a sphere and a rough half-space~\cite{pastewka2016contact}. An important conclusion of the both studies is that there's a linear regime of evolution of the contact area with the applied force. This regime, rather obvious from the physical point of view, extends over many orders of magnitude of the normal load. The added value of Pastewka and Robbins was identification of purely Hertzian contact regime at high loads, and an important extension to adhesive case, which is relevant for nano-metric contact systems. 

In our deterministic multi-asperity model with elastic interacting we can properly capture two contact regimes occurring at very light and at moderate loads. In the first regime~\cite{greenwood1970contact}, the contact occurs only at a single asperity and thus is accurately described by Hertzian theory of contact. The second regime is determined by the contact of multiple asperities, and results in linear evolution of the contact area as discussed in~\cite{greenwood1967elastic,pastewka2016contact}. Our model allows to accurately evaluate the transition between the two regimes, which can be easily predicted analytically. 
The transition load depends on the second and fourth spectral moments of the roughness and does not depend on the indenter's radius at least as long as it remains much bigger than asperity curvature. 
Contrary to the contribution of Pastewka and Robbins~\cite{pastewka2016contact}, we consider only light indentation depths of order of few surface's RMS, therefore we operate far from the purely Hertzian regime. That is why, the apparent contact radius is significantly bigger than the one predicted by the Hertzian contact, and the mean pressure, needed to estimate the nominal contact area cannot be properly evaluated. These results once again highlight the lack of scale separation between a macro-scale shape and surface roughness in non-conformal contacts, i.e. rigorously the effect of surface roughness cannot be simply included as an interfacial constitutive law in macroscopic simulations made with smooth surfaces.

The deterministic multi-asperity model can be easily extended to account for non-linear deformation of contacting asperities (viscous and plastic models can be used) while keeping the long-range interaction elastic. Adhesive contact of individual asperities also can be taken into account using either DMT~\cite{DMT}, JKR~\cite{JKR} or Maugis-Dugdale model~\cite{maugis1992adhesion,johnson1997adhesion}. The considered deterministic model does not inherit the main limitation of multi-asperity models, namely the absence of long-range elastic interactions between asperities. The coalescence of contacting asperities~\cite{greenwood2007w,afferrante2012interacting} in the considered limit of light loads apparently does not present a problem for the considered case, as for realistic root mean squared roughness, the contact area remains below $1$ \% in most considered cases.

Statistics of contact pressures clearly confirms the known fact that all contacting asperities operate in severe plastic regime. As was demonstrated here, including the size indentation effect does not change this conclusion. At light loads, the probability distribution of contact areas is found to follow exponential decay. In our model, the total pressure distribution for light pressures follows a power law with exponent $2/3$ and at higher loads decays as $\exp\left(-(p/p_0)^{3/2}\right)$, which is different from what Persson's model predicts for two nominally flat surfaces: linear growth of probability at  light pressures interval, and a Gaussian decay for higher pressures. Nevertheless, the demonstrated sub-linear growth of the pressure probability of relatively small pressure is in good agreement with recent results obtained with full deterministic numerical simulations~\cite{wang2017gauging}.

\vspace{1cm}
\noindent{\bf \Large Acknowledgment}\\[5pt]\normalsize

Helpful comments and remarks of Andrei Shvarts on the paper are greatly appreciated.

\appendix
\vspace{1cm}
\noindent{\bf \Large Appendix}
\normalsize

% \noindent{\bf \Large Appendix}\\[5pt]\normalsize

\section{Spectral density and asperity density\label{app}}

\begin{figure}[h!]
 \centering\includegraphics[width=1\textwidth]{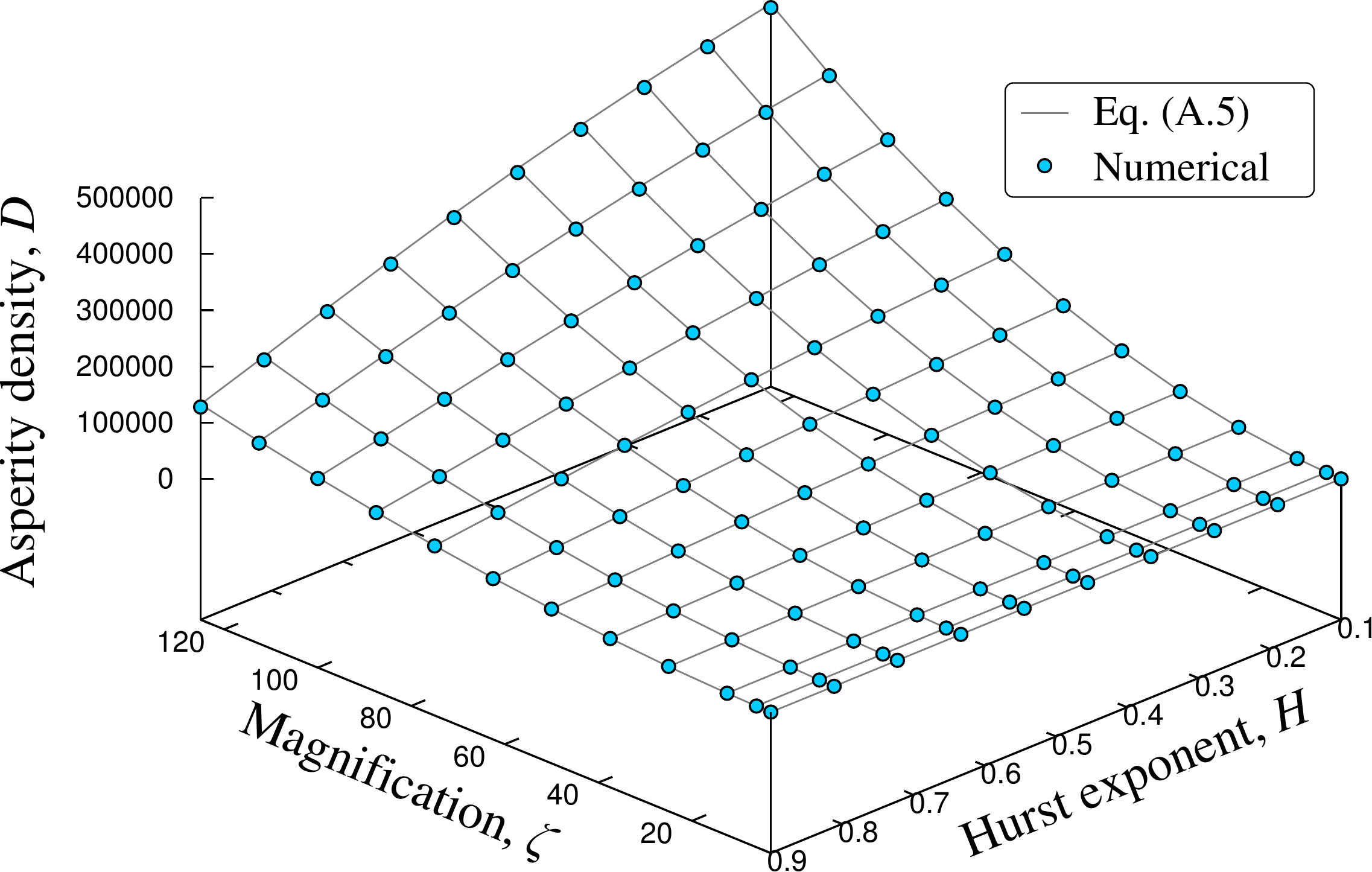}
 \caption{\label{fig:density}Comparison of Eq.~\eqref{eq:density} with numerically  evaluated asperity density of generated rough surfaces with $L/\lambda_l = 8$. Cross points of the analytical solution correspond to the same parameters used in numerical estimation.}
\end{figure}

Consider a surface spectrum $\Phi$ with a plateau which corresponds to an isotropic and periodic rough surface with a period $L$. Following the construction technique~\cite{hu1992ijmtm} for all possible directions determined by wavenumbers $k_x,k_y$ in $OX$ and $OY$ directions, respectively, in average for $K=\sqrt{k_x^2+k_y^2}$, the following equation holds:
 \begin{equation}
\langle \Phi(K) \rangle =\left\{ \begin{array}{ll}
	      \Phi_0 ,&\mbox{ if } \xi \le K/k_l \le 1\\
	      \Phi_0 (K/k_l)^{-2(1+H)},&\mbox{ if } 1 < K/k_l \le \zeta\\
	      0,&\mbox{ otherwise},
            \end{array}\right.
 \end{equation}
 where $k_l = 2\pi/\lambda_l$, $k_s=2\pi/\lambda_s$, $\xi = 2\pi/(Lk_l)$, $\zeta = \lambda_l/\lambda_s = k_s/k_l$, for the following inequalities $\lambda_s \le \lambda_l \le L$, where from it follows $\xi \le 1$, $\zeta \ge 1$.
A spectral moment $m_p$ is given by the following convolution
\begin{equation}
 m_{p} = \int\limits_{\frac{2\pi}{L}}^{k_s} K^{p+1} \Phi(K) dK \int\limits_{0}^{2\pi}\cos^{p}(\varphi)d\varphi  =
 \Phi_0T(p)\left[\left.\frac{K^{p+2}}{p+2}\right|_{\frac{2\pi}{L}}^{k_l} + \left.\frac{K^{p-2H}k_l^{2(1+H)}}{(p-2H)}\right|_{k_l}^{k_s}\right] 
\end{equation}
\begin{equation}
  m_{p} = \Phi_0 k_l^{p+2} T(p)\left[\frac{1-\xi^{p+2} }{p+2} + \frac{\zeta^{p-2H} - 1 }{p-2H}\right],
 \label{eq:a:mq0}
\end{equation}
where $T(p)$
$$
T(p) = \int\limits_{0}^{2\pi} \cos^p(\varphi)d\varphi,\quad T(0) = 2\pi, \quad T(2) = \pi,\quad T(4) = 3\pi/4.
$$
The density of asperities can be found as~\cite{nayak1971tasme} 
$$
  D = \frac{\sqrt3}{18\pi}\frac{m_4}{m_2},
$$
which by substituting expression for spectral moments~\eqref{eq:a:mq0} takes the following form~\cite{yastrebov2015ijss}:
\begin{equation}\displaystyle
D = \frac{\sqrt3 k_l^2}{24\pi} \left(\frac{1-\xi^{6}}{3} + \frac{\zeta^{4-2H} - 1 }{2-H}\right) \left/\left(\frac{1 - \xi^{4}}{2} + \frac{\zeta^{2-2H} - 1 }{1-H}\right)\right.,
\end{equation}
which for a surface without a plateau in the spectral density takes a simpler form
\begin{equation}\displaystyle
D \approx \frac{\sqrt3 k_l^2}{24\pi} \frac{1-H}{2-H}\frac{\zeta^{4-2H}}{\zeta^{2-2H} - 1 }.
\label{eq:density}
\end{equation}
Note that in the numerator we readily approximated $(\zeta^{4-2H}-1)$ by $\zeta^{4-2H}$; however, this approximation cannot be made in the denominator if $H$ is reasonably high. In Fig.~\ref{fig:density} we compare this analytical expression~\eqref{eq:density} with numerically computed asperity densities for surfaces with different Hurst exponents $H \in [0.1,0.9]$ and different magnifications $\zeta = k_s/k_l \in [3.125, 125]$, to keep the surfaces representative we fixed the longest wavelength to $L/\lambda_l = 8$ and considered surfaces without a plateau. Note that to ensure accuracy of asperity detection, for $L/\lambda \le 500$ ($\zeta \le 62.5$) we used surfaces with discretization $N = 4096$, and for $500 < L/\lambda_s \le 1000$ ($62.5  < \zeta \le  125$), discretization $N = 8192$ was used. This choice was made because a considerable underestimation of asperity densities was observed for insufficient surface discretization, e.g. the relative error for $N=4096$, $L/\lambda_l = 8$, $L/\lambda_s = 1000$, reached $\approx 8$\% for $H=0.9$ and raised to $\approx 15\%$ for $H=0.1$.

The average distance between closest asperities for a surface without plateau is given by:
$$
  \langle d\rangle = 1/\sqrt{D} \approx \sqrt{\frac{24\pi}{\sqrt3 k_l^2} \frac{(2-H)}{(1-H)}\frac{(\zeta^{2-2H} - 1)}{\zeta^{4-2H}}} = \lambda_l \sqrt{\frac{6 }{\sqrt3  \pi } \frac{(2-H)}{(1-H)}\frac{(\zeta^{2-2H} - 1)}{\zeta^{4-2H}}}. 
$$ 
As close asperities have a comparable height, the critical contact radius for asperities, 
at which the associated contact zones coalesce, may be estimated as $a' = \langle d\rangle/2$. Moreover, at average contact radius $a_c = \langle d\rangle/4$ the underlying surface would change curvature and thus the Hertzian assumption cannot be valid anymore. Therefore we shall assume that that the critical contact radius for asperities, beyond which the solution is not valid, is given by
\begin{equation}
  a_c = \lambda_l \sqrt{\frac{\sqrt3 }{8  \pi } \frac{(2-H)}{(1-H)}\frac{(\zeta^{2-2H} - 1)}{\zeta^{4-2H}}}.
\end{equation}

% \bibliographystyle{abbrv} 
% \bibliography{references}

\end{document}